\documentclass[english,reprint,amsmath,amssymb,aps,pra,superscriptaddress,floatfix]{revtex4-1}

\usepackage{dcolumn}
\usepackage{bm}

\usepackage{tabularx}
\usepackage{graphicx}
\usepackage{babel}
\usepackage{layouts}
\usepackage{epstopdf}
\usepackage{physics}

\usepackage{dcolumn}
\usepackage{bm}
\usepackage{amsmath}
\usepackage{comment}
\usepackage{bbold}
\DeclareMathOperator{\di}{d\!}

\usepackage[unicode=true,pdfusetitle,
 bookmarks=true,bookmarksnumbered=false,bookmarksopen=false,
 breaklinks=false,pdfborder={0 0 0},backref=false,colorlinks=true]
 {hyperref}
 
 \usepackage[normalem]{ulem}

\usepackage{color}

\makeatletter
\@ifundefined{textcolor}{}
{%
 \definecolor{BLACK}{gray}{0}
 \definecolor{WHITE}{gray}{1}
 \definecolor{RED}{rgb}{1,0,0}
 \definecolor{GREEN}{rgb}{0,1,0}
 \definecolor{BLUE}{rgb}{0,0,1}
 \definecolor{CYAN}{cmyk}{1,0,0,0}
 \definecolor{MAGENTA}{cmyk}{0,1,0,0}
 \definecolor{YELLOW}{cmyk}{0,0,1,0}
 }

\newcommand{\bit}{\begin{itemize}}
\newcommand{\eit}{\end{itemize}}

\newcommand{\bea}{\begin{eqnarray}}
\newcommand{\eea}{\end{eqnarray}}

\newcommand{\be}{\begin{equation}}
\newcommand{\ee}{\end{equation}}

\newcommand{\sgn}{\,\mbox{sgn}}
\newcommand{\B}{\mathbf}

\definecolor{dgreen}{rgb}{0.0, 0.5, 0.0}

\makeatletter

\begin{document}

\title{Transport of neutral optical excitations using electric fields}

\author{Ovidiu Cotle\c{t}}
\thanks{These authors have contributed equally to this work.}
\affiliation{Institute of Quantum Electronics, ETH Z{\"u}rich, CH-8093, Z\"urich, Switzerland }

\author{Falko Pientka}
\thanks{These authors have contributed equally to this work.}
 \affiliation{Department of Physics, Harvard University, Cambridge, Massachusetts 02138, USA}

\author{Richard Schmidt}
 \affiliation{Department of Physics, Harvard University, Cambridge, Massachusetts 02138, USA}
\affiliation{Max Planck Institute of Quantum Optics,  85748 Garching, Germany}

\author{Gergely Zarand}
\affiliation{Department of Theoretical Physics,  Institute of Physics,  Budapest University of Technology and Economics, H-1521, Hungary}

 \author{Eugene Demler}
\affiliation{Department of Physics, Harvard University, Cambridge, Massachusetts 02138, USA}

\author{Atac Imamoglu}
\affiliation{Institute of Quantum Electronics, ETH Z{\"u}rich, CH-8093, Z\"urich, Switzerland }

\begin{abstract}
{Mobile quantum impurities interacting with a fermionic bath form quasiparticles known as Fermi polarons. We demonstrate that a force applied to the bath particles can generate a drag force of similar magnitude acting on the impurities, realizing a novel, nonperturbative Coulomb drag effect. To prove this, we calculate the fully self-consistent, frequency-dependent transconductivity at zero temperature in the Baym-Kadanoff conserving approximation.
We apply our theory to excitons and exciton polaritons interacting with a bath of charge carriers in a doped semiconductor embedded in a microcavity. In external electric and magnetic fields, the drag effect enables electrical control of excitons and may pave the way for the implementation of gauge fields for excitons and polaritons. Moreover, a reciprocal effect may facilitate optical manipulation of electron transport. Our findings establish transport measurements as a novel, powerful tool for probing the many-body physics of mobile quantum impurities.
}
\end{abstract}

\maketitle

\section{Introduction}
Polaritons are composite bosonic particles formed by hybridization of propagating photons and quanta of polarization waves in a solid. A particular realization that has recently been extensively studied involves two-dimensional (2D) cavity exciton polaritons, implemented in monolithic III-V semiconductor heterostructures \cite{weisbuch1992observation}, as well as in transition metal dichalcogenide (TMD) monolayers embedded in open dielectric cavities \cite{liu2015strong,dufferwiel2017valley,sidler2017fermi}. Remarkably, these cavity-polariton excitations combine an ultra-light effective mass dictated by their photonic content with a sizable interparticle interaction strength stemming from their excitonic character. This unique combination allows for the realization of a driven-dissipative interacting boson system that has been shown to exhibit a myriad of many-body phenomena such as nonequilibrium condensation \cite{imamog1996nonequilibrium,baumberg2000parametric,carusotto2005spontaneous,kasprzak2006bose,deng2010exciton}, superfluidity \cite{amo2009superfluidity} and the Josephson effect \cite{abbarchi2013macroscopic}.

Despite this progress, the realization of topological states of polaritons remains elusive and  finding means to implement the required effective gauge fields is a challenge for theory and experiment. Exciton polaritons are charge-neutral particles and, hence, it is not possible to couple them directly to dc electric or magnetic fields. Instead, it might be  possible to exploit their interaction with charge carriers to  mediate such a coupling. In this regard semiconducting TMD monolayers are a particularly promising platform featuring exceptionally large exciton binding energies and strongly attractive interactions between excitons and charge carriers.

In a simple picture, excitons in the presence of denerate electrons can be considered as mobile impurities interacting with a Fermi sea, which constitutes a fundamental problem of many-body physics \cite{Josephson1969,Doniach1970,Kondo1983,Calleja1991,Kagan1992,rosch1999quantum,prok2008,schmidt2012fermi,schmidt2017,massignan2014polarons}. To lower its energy, an exciton can bind an additional charge carrier forming a charged trion. Alternatively, the exciton can, however, also create a polarization cloud in its environment forming an attractive exciton polaron, shortly referred to as polaron. In this case the exciton does remain a neutral particle, that is dressed by fluctuations of the Fermi sea, which renormalize its energy and effective mass. The competition of trion  formation and polaronic dressing was previously observed in cold atomic systems near a Feshbach resonance in two and three dimensions \cite{schirotzek2009observation,koschorreck2012attractive,Zhang2012,kohstall_metastability_2012}.  Recently, it was demonstrated both theoretically \cite{sidler2017fermi,efimkin2017many} and experimentally \cite{sidler2017fermi,back2017giant} that polaron physics also plays a key role in semiconductor photonic materials hosting a two dimensional electron system (2DES). In particular, absorption experiments in TMD monolayers show one of the key signatures of Fermi polaron formation: a red-shifted optical resonance with an oscillator strength that increases with increasing electron density $n_e$, which is accompanied by a strong blue-shift and broadening of the bare-exciton resonance \cite{back2017giant}. These observations demonstrate that the optical excitation spectrum should be described in terms of attractive and repulsive exciton polarons. In contrast, trion states have negligible weight in absorption spectra due to their vanishing oscillator strength.

In this work, we propose to use exciton polarons as a new means to control photons by dc electric or magnetic fields. In contrast to trions, polarons are amenable to polariton formation and recent TMD-cavity experiments have demonstrated strong coupling of exciton polarons and light \cite{sidler2017fermi}. We find that, although polarons are charge-neutral optical excitations, their interaction with the Fermi sea forces them to follow the motion of charge carriers in an electric or magnetic field. This phenomenon bears similarity to Coulomb drag between two semiconductor layers \cite{Zheng1993,Kamenev1995,Flensberg1995,narozhny2016coulomb}, motivating us to coin the term polaron drag. 
A key difference to conventional drag \cite{berman2010drag}, however, originates from the nonperturbative nature of the polaronic coupling, which gives rise to a remarkably efficient drag mechanism. Indeed, we find that the zero-temperature drag conductivity of polarons can ideally be of the same order as the electron conductivity, promising the realization of sizable photonic gauge fields.
From a more general perspective, we describe a novel transport mechanism that is mediated by interactions beyond the perturbative regime. This paves the way to probing quantum many-body effects using transport measurements. Our results apply generically to Fermi polaron systems at low density in cold atom or condensed matter settings.

We start our analysis by a heuristic derivation of the polaron drag force based on intuitive arguments in Sec.~\ref{sec:heuristic}. Moreover, we solve the semiclassical equations of motion of a polaron in an electric. In Sec.~\ref{sec:diagrammatics} we corroborate these results by a microscopic calculation of the polaron drag conductivity in an electric field using diagrammatic perturbation theory within the Kadanoff-Baym conserving approximation. To make this paper self-contained, we start this section by reviewing the Fermi polaron problem as well as the conserving approximation.  In Sec.~\ref{sec:polaritons} we extend our results to polaritons and comment on nonequilibrium effects in Sec.~\ref{sec:nonequilibrium}. After discussing extension of our analysis to the case of magnetic field response in Sec.~\ref{sec:magneticresponse}, we conclude by presenting an outlook on possible extensions of our work.

\section{Heuristic derivation of polaron drag}\label{sec:heuristic}

Let us consider a mobile exciton with mass $m_x$ interacting with a Fermi sea of electrons with mass $m_e$. Applying a force on the electrons, causes an acceleration ${\bf a}_e$ of the Fermi sea. We now go to the rest frame of the electrons. Because this is a non-inertial reference frame, a fictitious force acts on the exciton polaron ${\bf F}=-m_x {\bf a}_e$ accelerating them. Crucially the force is proportional to the bare mass of the exciton. The interaction with the electrons, however, impairs the motion of the polaron and, hence, the polaron is heavier than the bare exciton with an effective mass $m_x^*$. The acceleration of the polaron in the electron rest frame is thus ${\bf F}/m_x^*=-{\bf a}_e m_x/m_x^*$. Going back to the lab frame, we have to add the acceleration of the reference frame and we thus arrive at an exciton-polaron acceleration
\begin{align}
 {\bf a}_x=\Bigl(1-\frac{m_x}{m_x^*}\Bigr){\bf a}_e\label{drag_acc}
\end{align}
This equation relates any force acting on the electron system to a somewhat smaller force on the polarons.
This force can be qualitatively understood as friction between electrons and excitons, which originates from the ability of the excitons to minimize their energy by following the polarization cloud surrounding them.
This result has several important ramifications:

(i) If the polaron mass significantly exceeds the bare exciton mass, polaron drag can be extremely efficient promising a drag mobility of the same order as the electron mobility. This is a consequence of the nonperturbative interaction that is responsible for the polaron formation.

(ii) Polaron drag is present even at zero temperature. This is in stark contrast to conventional Coulomb drag in semiconductor bilayers, where a phase space argument predicts a $T^2$ behavior.

(iii) All forces acting on electrons translate to polarons according to Eq.~(\ref{drag_acc}). Hence polarons also experience a Hall effect in a magnetic field and they can scatter off electronic disorder.

(iv) The drag force acts in the same direction for both repulsive and attractive polarons. The direction of motion in an electric field only depends on the charge of the carriers in the Fermi sea.

(v) In the limit of weak interactions $m_x^*\to m_x$ and the polaron drag becomes inefficient as expected in the perturbative regime.

(iv) The electrons should also experience a drag force, when the excitons are accelerated by an external force. Equation~(\ref{drag_acc}) corresponds to the combined force all electrons exert on a single polaron.
Hence, the average inverse drag force on an electron in the Fermi sea in the presence of an exciton force $\B F_x$ should be $(n_x/n_e)(1-m_x/m_x^*) \B F_x$, where $n_{e,x}$ is the density of electrons (excitons).

With this result we can write down general semiclassical equations of motion:
\begin{align}
m_e\frac{\di }{\di t } \B v_e(t) &= \B F_e(t)+\frac{n_x}{n_e} \left(m_x^*-m_x \right) \frac{\B F_x(t)}{m_x^*}\label{eom_el}\\
m_x^*\frac{\di }{\di t } \B v_{x}(t) &=\B F_x(t) +\left(m_x^*-m_x\right) \frac{\B F_e(t)}{m_e}\label{eom_x}
\end{align}
We can evaluate the polaron drag conductivity in an ac electric field $\B E(t)$ in the presence of disorder for both electrons and excitons by choosing
\begin{align}
\B F_e(t)&= -e \B E(t) -\frac{m_e \B v_e(t)}{\tau_e},
&\B F_x(t) = -\frac{m_x^* \B v_x(t)}{\tau_x^*}\label{eom_x_dis},
\end{align}
where $\tau_{e}$($\tau_x^*$) is the transport lifetime of electrons (exciton-polarons). To leading order in the polaron density, we can ignore the drag force on electrons. The solution in Fourier space reads
\begin{align}
\B v_e(\Omega) &= \frac{-e \B E(\Omega)}{m_e}\frac{\tau_e}{1-i \Omega \tau_e} \label{eq2}\\
\B v_{x}(\Omega) &=\Bigl(1 -\frac{m_x}{m_x^*}\Bigr)\Bigl(\frac{-i \Omega \tau_x^* }{1-i \Omega \tau_x^*}\Bigr)\B v_e(\Omega) \label{v_x}.
\end{align}
Notice that in the absence of excitonic disorder, $\tau_x^*\to\infty$, the drag is the only force acting on the excitons and their drift velocity is simply $v_x=(1-m_x/m_x^*)v_e$. This case is particularly relevant for exciton polaritons, which are expected to be largely immune to exciton disorder. We emphasize that this results only holds at small polaron densities as we neglect electron drag forces generated by polarons. From the exciton drift velocity, we obtain the transconductivity to first order in the polaron density $n_x$
\begin{align}
\sigma_{xe}(\Omega)= \frac{e n_{x}}{m_e}\left(1 -\frac{m_x}{m_x^*}\right)\frac{i \Omega \tau_e \tau_x^* }{\left(1-i \Omega \tau_e\right)\left(1-i \Omega \tau_x^*\right)} .\label{transconductivity}
\end{align}
It is important to note that our analysis so far has neglected incoherent scattering of electrons and polarons.  In addition to coherent scattering of electrons and excitons that lead to polaron formation there may also be incoherent scattering events that lead to a finite lifetime of polarons in an excited state. We investigate this effect in more detail in Sec.~\ref{sec:calculate_conductivity} and \ref{sec:Fdis}, and discuss in what parameter regimes it can be neglected.  We comment on nonequilibrium effects, which may become important in experiments, in Sec.~\ref{sec:polaritons}.

As mentioned above, the reverse effect also exists: an electric current should flow in response to an ac force $\B f_x(t)$ applied to the excitons. Such a force can be effected by applying an ac field perpendicular to the 2D plane that modifies the exciton energy through a quantum-confined Stark effect with a spatial and/or time dependence determined through the applied laser field \cite{sie2017large}.   We can find the drag conductivity of electrons from Eqs.~(\ref{eom_el}) and (\ref{eom_x}) with
\begin{align}
\B F_e(t)&= -\frac{m_e \B v_e(t)}{\tau_e},
&\B F_x(t) =\B f_x(t) -\frac{m_x^* \B v_x(t)}{\tau_x^*}.
\end{align}
In this case, we need to include the drag force on electrons in Eq.~(\ref{eom_el}). We can, however, neglect the drag term in the polaron equation of motion (\ref{eom_x}), since $F_e\propto v_e\propto n_x/n_e$ only contributes at higher order in polaron density.
With this we again arrive at a transconductivity given by Eq.~(\ref{transconductivity}), as guaranteed by Onsager's reciprocity principle. An experimentally more relevant quantity is the electric voltage which builds up as a response to an exciton force when no current can flow. The electric field can be found by setting $\B v_e=0$ in the equations of motion which yields
\begin{align}
e\B E(\Omega)=\frac{n_x}{n_e} \Bigl(1 -\frac{m_x}{m_x^*}\Bigr)\Bigl(\frac{-i \Omega \tau_x^* }{1-i \Omega \tau_x^*}\Bigr) \B f_x(\Omega)
\end{align}
We mention in passing that the semiclassical analysis presented here can be extended to include an external magnetic field (see also \cite{Efimkin2017,ravets2018polaron}). The drag force then gives rise to a Hall effect and cyclotron resonances of exciton polarons. We defer a discussion of these phenomena to Sec.~\ref{sec:magneticresponse} and instead proceed with a diagrammatic calculation of the transconductivity, which corroborates the main findings of this section.

\section{Diagrammatic calculation of transconductivity using the conserving approximation}\label{sec:diagrammatics}

We now evaluate polaron drag within a microscopic theory in order to verify the heuristic results discussed in the previous section. To this end, we use diagrammatic perturbation theory taking into account the effect of electron-exciton interactions as well as disorder for electrons and excitons. We are interested in a nonperturbative effect of the interaction and must therefore proceed with care. Simply evaluating a certain class of diagrams might lead to erroneous results as an incomplete set of diagrams does not necessarily satisfy the conservation laws of the physical systems. A powerful technique to generate diagrams obeying conservation laws is the {\em conserving approximation} (see \cite{Stefanucci2013} for a pedagogical introduction).

We start by introducing the Hamiltonian of the exciton electron system and discussing the polaron problem in the absence of fields. We then review the basic principles of the conserving approximation and use it to find an approximation to the drag conductivity within linear response.
As a crucial simplification, we focus on the limit of small polaron density. This limit is also implicitly assumed in the semiclassical analysis in Sec.~\ref{sec:heuristic} as we neglect interactions between polarons. Moreover the quasiparticle picture of polarons eventually breaks down at sufficiently high densities. A small polaron density also justifies the standard diagrammatic description of polarons in the ladder approximation presented in Sec.~\ref{sec:polaron_problem}. Hence, our calculation will only include contributions to leading order in the polaron density.

\subsection{Exciton polaron problem}\label{sec:polaron_problem}

We consider noninteracting excitons coupled via contact interactions of strength $V$ to a Fermi sea of noninteracting electrons. To be specific we focus on the two-dimensional case, however, our results for the transconductivity apply to three dimensions as well. We focus on the limit of very small exciton density where the statistics of excitons becomes irrelevant. We exploit this fact by treating the low-density excitons as an effective Fermi gas which allows to simplify calculations. Since we are moreover interested in the regime where no exciton condensation takes place, treating excitons effectively as fermions allows to take the corresponding limit of $T\to 0$ without technical complications from Bose condensation. We emphasize that all results are independent of the statistics.

Our system is described by the Hamiltonian
\begin{widetext}
\begin{align}
H=&\int \di \B r \left\{ \Psi^\dagger_e(\B r)\left(-\frac{\nabla^2}{2 m_e}-\mu_e \right)\Psi_e(\B r)+ \Psi_x^\dagger(\B r)\left(-\frac{\nabla^2}{2 m_x}  - \mu_x \right) \Psi_x(\B r)+ V \Psi_x^\dagger(\B r) \Psi_x(\B r) \Psi_e^\dagger(\B r) \Psi_e (\B r) \right. \label{hamil}  \\ 
&+ \left.  \Psi_e^\dagger (\B r) \Psi_e(\B r) U^{(e)} (\B r ) + \Psi_x^\dagger (\B r) \Psi_x(\B r) U^{(x)} (\B r )  \right\}\nonumber
\end{align}
\end{widetext}
where we introduced the creation operators of the electrons and excitons $\Psi_e^\dagger$ and $\Psi_x^\dagger$ and the chemical potentials $\mu_{e,x}$. We model the disorder potentials $U^{(e,x)}$ by Gaussian ensembles with zero means and variances $\langle U^{(e,x)}(\B r)U^{(e,x)}(\B r')\rangle=\gamma_{e,x}\delta(\B r -\B r')$ where $\gamma$ parameterizes the strength of disorder.

\begin{figure}
\includegraphics[width=\columnwidth]{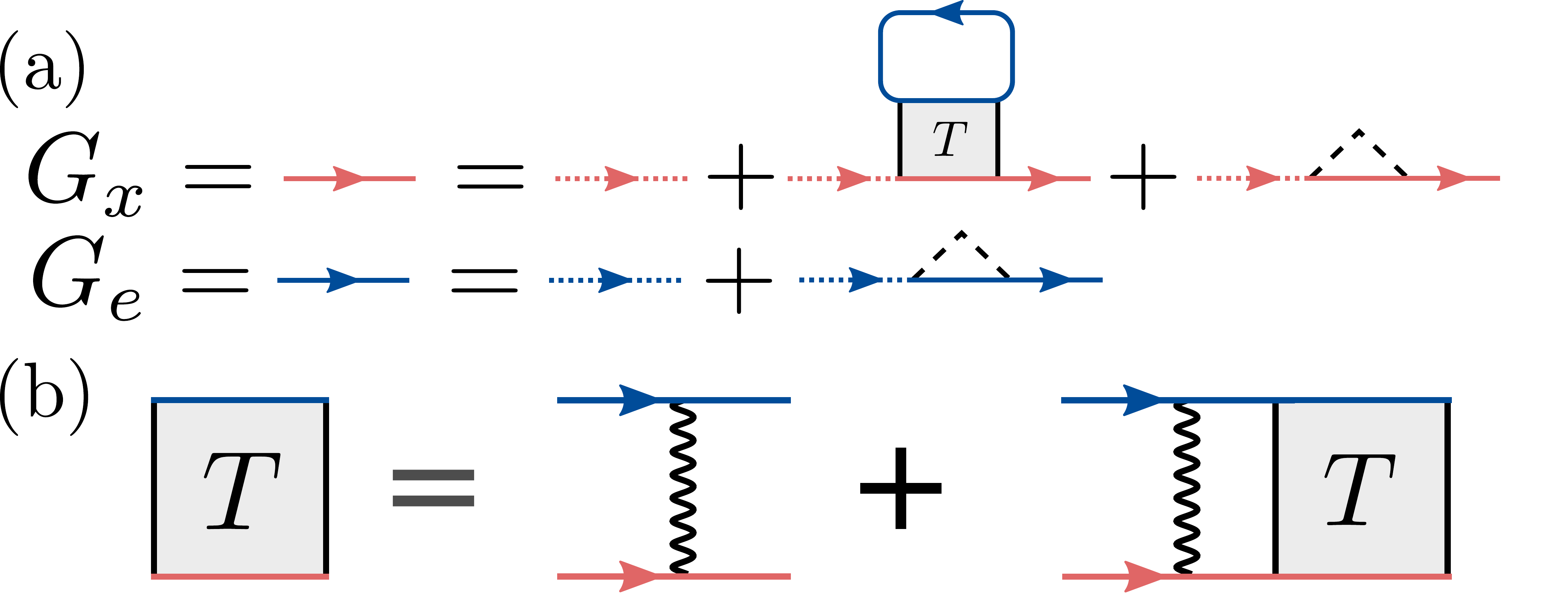}
\caption{The Green's functions and T matrix.}
\label{fig:Tmatrix}
\end{figure}

We introduce the exciton and electron Green's functions
\begin{align}
G_{e,x}(\B r' t'; \B r , t ) = - i \langle T \Psi_{e,x}^\dagger(\B r',t') \Psi_{e,x}(\B r,t )  \rangle
\end{align}
After disorder averaging the system is translationally invariant, which allows us to work in Fourier space
\begin{align}
G_{e,x}(\B r, t) \equiv \int dp G_{e,x}(p) e^{i \B p \B r - i \omega t} ,
\end{align}
where we have introduced $p\equiv\left(\B p,\omega \right)$ for notational simplicity [we will later also use $k\equiv (\B k,\epsilon)$]. Moreover, we incorporate factors of $1/2 \pi$ into the definition of integral measure such that $\di p=d \B p \di \omega/(2\pi)^{3}$. The effect of disorder and interactions is taken into account by introducing self energy corrections to the Green's functions shown in Fig.~\ref{fig:Tmatrix}(a)
\begin{align}
 G_e^{-1} (p) &= \omega-\xi_\B p - \Sigma_{{\rm dis},e}(p)   \label{Ge}\\
 G_x^{-1} (p) &= \omega-\omega_\B p - \Sigma_{\rm int}(p) -\Sigma_{{\rm dis},x}(p)  \label{Gx}
\end{align}
where we introduced the dispersions of electrons $\xi_{\B p}=\B p^2/(2 m_e)-\mu_e$ and excitons $\omega_{\B p}=\B p^2 /(2 m_x)-\mu_x$ and the exciton self energy from interactions with electrons $\Sigma_{{\rm int}}$ as well as the disorder self energies $\Sigma_{e,{\rm dis}}$ and $\Sigma_{x,{\rm dis}}$. Assuming a small density of polarons in the system, we can neglect the effect of excitons on the electron system.

The calculation of the self energies requires some approximations. We treat the effect of disorder in the self-consistent Born approximation and therefore ignore any quantum interference effects such as weak-localization. For the evaluation of the interaction self energy $\Sigma_{{\rm int}}$, we use the self-consistent T matrix approach in the ladder approximation displayed in Fig.~\ref{fig:Tmatrix}(b). This choice gives the leading contribution in the limit of small exciton density \cite{Galitskii1958} and has been successfully used to describe the physics of Feshbach resonances in cold atomic systems \cite{prok2008,schmidt_excitation_2011}. The T matrix does not have any vertex corrections due to disorder because of our choice of Gaussian correlated white-noise disorder.
We will see in Sec.~\ref{sec:conserving}, how these self energies can be obtained within a conserving approximation.  We point out that the non-self-consistent T matrix approximation is equivalent to the Chevy ansatz for the variational polaron wavefunction \cite{chevy2006universal,Lobo2006}.

The disorder self energies can be evaluated as
\begin{align}
\Sigma_{{\rm dis},e/x}(p)=\int   \frac{d \B k}{(2\pi)^2}  \gamma_{e/x}  G_{e/x}(k)|_{\epsilon=\omega} =\frac{-i }{2 \tau_{e/x}}\mathrm{sgn}(\omega) .\label{dis}
\end{align}
where the lifetimes are defined as
\begin{align}
\tau_{e,x}=\frac{1}{2\pi\rho_{e,x}\gamma_{e,x}}\label{tau_ex}
\end{align}
with $\rho_{e,x}$ the densities of states per unit volume at the Fermi surface. Importantly, in the self-consistent theory the exciton Green's functions in Eq.~(\ref{dis}) is dressed by disorder and interactions and hence $\rho_x$ refers to the renormalized exciton dispersion to be determined below.

The exciton self energy due to interactions can be evaluated in the self-consistent T-matrix approximation
\begin{align}
\Sigma_{\rm int}(p) =&  -i \int \di k G_e(k) T(k+p)\label{selfenergy}
\end{align}
where we introduced the self-consistent T-matrix shown in Fig.~\ref{fig:Tmatrix}(b)
\begin{align}
 T(p)=&V+i V \int \di k G_e(k) G_x(p-k) T(p).
\end{align}
The contact interaction allows for a simple solution
\begin{align}
\label{T1}
T^{-1}(p)=V^{-1}- i \int \di k G_e(k) G_x(p-k) .
\end{align}
In two dimensions, a bound state of electrons and excitons (i.e., a trion) exists for arbitrarily weak interactions in the limit of a single exciton, and its energy $\epsilon_T$ is determined by the pole of the T matrix. 

\begin{figure}
\includegraphics[width= \columnwidth]{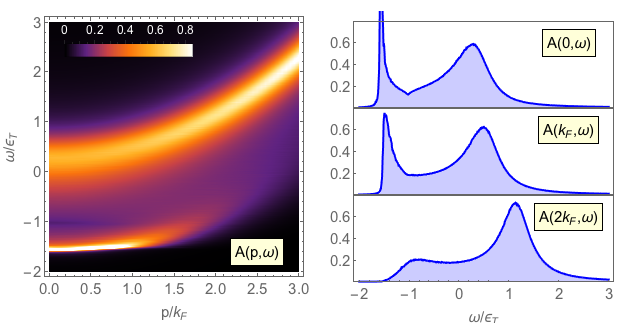}
\caption{Self-consistent spectral function of excitons at zero exciton density, $\mu_e = \epsilon_T/2$, $m_x=2m_e$, and disorder broadening $1/2\tau_x=\epsilon_T/100$.}
\label{fig:SF}
\end{figure}

At vanishing exciton density (i.e. $n_x=0$) we can evaluate the self energy by replacing $G_x\to G_x^R$ in the above equations (for more details about this step see Sec.~\ref{sec:low_density}). We can evaluate all frequency integrals by closing the contour in the upper half-plane, where the retarded exciton Green's function is analytical, and obtain 
\begin{align}
 \Sigma^{(0)}_{\rm int}(p)& =   \int \frac{\di\B k }{(2 \pi)^2} n_F(\xi_\B k)  T^{(0)}(\omega+\xi_\B k,\B k+\B p)\label{eq20},\\
T^{(0)}(p)^{-1} &=V^{-1} + \int \frac{\di \B k}{(2 \pi)^2}[1-n_F(\xi_\B k)]\notag\\
&\qquad \qquad \qquad \times G_x^R(\omega-\xi_\B k,\B p-\B k),\label{eq21}
\end{align}
where the superscript $^{(0)}$ denotes quantities at $n_x=0$ and $n_F(x)$ is the Fermi-Dirac distribution function. In order to regularize the contact interaction $V$, we introduce a UV momentum cutoff $\Lambda$. The interaction strength is then related to the experimentally accessible trion binding energy $\epsilon_T$ at zero electron density by \cite{schmidt2012fermi}
\begin{align}
 V^{-1}=-\int_{|\B k|<\Lambda} \frac{\di \B k}{(2 \pi)^2} \frac{1}{\epsilon_T + \frac{\B k^2}{2 m_e}+\frac{\B k^2}{2 m_x}}.
\end{align}
We solve Eq.~(\ref{eq20}) self-consistently, by discretizing momentum and energy and using an iterative method. The self-consistent exciton spectral function $A(p)=- \pi^{-1} {\rm Im}G^R_x(p)$ for $n_x=0$ and $\mu_e=\epsilon_T/2$ is plotted in Fig.~(\ref{fig:SF}).

The plot shows two spectral features: at negative frequencies the \textit{attractive polaron} is a well-defined quasiparticle excitation at sufficiently low momenta, while a second well-defined, metastable \textit{repulsive} polaron quasiparticle exists at positive energies. Both excitations have been observed in transition metal dichalcogenides \cite{sidler2017fermi} and cold atomic quantum gases close to Feshbach resonances \cite{koschorreck2012attractive,kohstall_metastability_2012,cetina_2016,Scazza2016,schmidt2017}.

\subsubsection{Attractive polaron quasiparticles}

The ground state of a single exciton described by the model in Eq.~(\ref{hamil}) depends on the dimensionless interaction strength given by the ratio $\epsilon_F/\epsilon_T$ of Fermi energy of electrons and trion energy. While the attractive polaron is stable at higher electron densities, diagrammatic Monte Carlo simulations predict a trion ground state for Fermi energies below $ 0.1 \epsilon_T$ for contact interaction models \cite{Vlietinck2014,Kroiss2014}. We henceforth assume a sufficiently large electron density so that the physics at low exciton densities is dominated by the formation of attractive polarons. In this regime, it is instructive to introduce an effective (or projected) Green's function $\bar{G}_x$, describing the propagation of attractive polaron quasiparticles (see App.~\ref{app:effective}):
\bea
\bar{G}_x(p) = \frac{1}{\omega-\zeta_\B p+i/2\tau_x^*(\B p) \sgn (\omega) }\label{Gatt}.
\eea
Here we have introduced the polaron dispersion 
\begin{align}
\zeta_\B p=\frac{\B p^2}{2m_x^*}-\mu_x^* 
\end{align}
with the polaron chemical potential $\mu^*_x$ measured from the bottom of the polaron band. Moreover, we have defined the effective polaron mass $m^*_x$, the polaron lifetime $\tau^*_x(\B p)$, and the quasiparticle weight $Z$ as
\begin{align}
\left(m_x^*\right)^{-1}=&Zm_x^{-1} +Z\partial^2_{\B p}\mathrm{Re}\Sigma_{\rm int}(\B p,0)\Bigr|_{\B p=\B p_F}\label{self_deriv1}\\
1/2\tau_x^*(\B p)=&Z /2 \tau_x - Z {\rm Im}\Sigma_{\rm int}(\B p,\zeta_\B p))\label{lifetime}\\
Z^{-1}=&1-\partial_\omega  \Sigma_{\rm int}(\B p_F,\omega)\Bigr|_{\omega=0}\label{self_deriv2}.
\end{align}
where we introduced the exciton Fermi momentum $p_F$. The polaron density of states in Eq.~(\ref{tau_ex}) is hence $\rho_x=m^*_x/2\pi$. The polaron lifetime has a constant part from disorder scattering as well as a momentum dependent part due to incoherent electron-polaron scattering (see App.~\ref{sec:scattering} for an estimate of ${\rm Im}\Sigma_{\rm int}(p)$ as well as the discussion in Ref.~\cite{schmidt_excitation_2011}). We emphasize that it is crucial to evaluate the polaron self energy self-consistently to obtain the momentum-dependent lifetime. 

In Sec.~\ref{sec:calculate_conductivity}, we shall make use of the fact that the full exciton Green's function can be approximated by the effective expression, $G_x(p)\simeq Z\bar{G}_x(p)$ near the resonance $\omega\simeq \zeta_\B p$ and $|\B p|\ll k_F$. The interaction vertex between electrons and polarons, however, is determined by virtual transitions to excited states and knowledge of the full exciton Green's function is required to accurately determine the vertex functions.

\subsection{Conserving approximation}
\label{sec:conserving}

We now turn to finding an expression for the transconductivity within the conserving approximation \cite{Baym1961,Stefanucci2013}. We first review the basic principles of the conserving approximation and then apply this formalism to the polaron problem. In this section we temporarily adopt the Keldysh notation, which is most convenient for this purpose.

Intuitively, an approximation that satisfies conservation laws can be derived from a quantity that is invariant under symmetry operations. In the diagrammatic language such quantities are represented by vacuum diagrams, i.e., diagrams that appear in the expansion of the thermodynamic potential.
Our starting point is a functional $\Phi[G]$ of the Green's function defined as the sum of certain two-particle irreducible connected vacuum diagrams. The choice of included diagrams determines the accuracy of the approximation.

We can obtain a self energy from the generating functional $\Phi[G]$ by a functional derivative
\begin{align}
\Sigma(1,2)=\frac{\delta \Phi[G]}{\delta G(2,1)},
\end{align}
where the arguments are space-time coordinates on the Keldysh contour. In a diagrammatic language, this procedure amounts to simply cutting a Green's function line in the vacuum diagrams leading to the desired self energy, which is called $\Phi${\em -derivable}.
Crucially, the Green's function obtained from a Dyson equation with this self energy turns out to be conserving, i.e., physical quantities constructed with this Green's function obey conservation laws such as the continuity equation \cite{Baym1962}. Importantly, $\Sigma$ needs to be evaluated self-consistently, which means that all internal Green's function lines in $\Sigma$ represent full lines dressed by the self energy.

\begin{figure}
\includegraphics[width=\columnwidth]{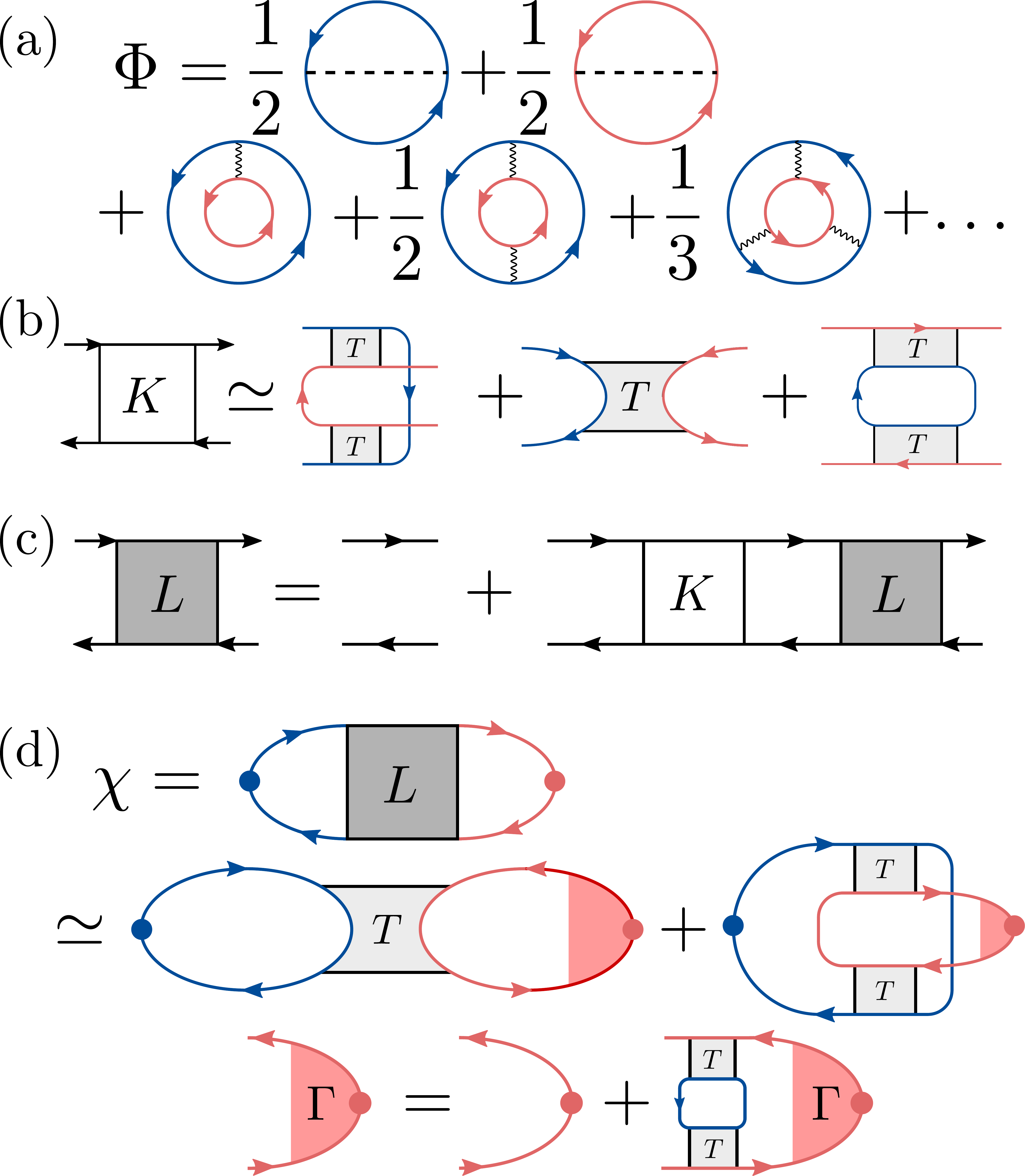}
\caption{Linear response theory for polaron drag within the conserving approximation to lowest order in polaron density. (a) The functional $\Phi$. Dashed (wavy) lines represent disorder (interactions). Blue (dark gray) lines indicate dressed electron propagators and red (light gray) lines indicate dressed exciton propagators as defined in Fig.~\ref{fig:Tmatrix}.
(b) The irreducible two particle vertex $K=\delta^2\Phi/\delta G^2$. Only diagrams to leading order in exciton density are retained. (c) The Bethe-Salpeter equation (\ref{bse}) for the reducible two-particle vertex $L$.
(d) The transconductivity diagram obtained from Eq.~(\ref{response_function}). The second and third line show the solution for $L$ based on the irreducible vertex $K$ in panel (b).}
\label{fig:conserving}
\end{figure}

Response functions can also be derived within the conserving approximation (for a detailed discussion see \cite{Stefanucci2013}). We start from the equation of motion
\begin{align}
 \int d2 G^{-1}(1,2)G(2,3)=\delta(1,3)\label{eom}
\end{align}
where we have defined the operator
\begin{align}
G^{-1}(1,2)=[i\partial_\tau- h(1)]\delta(1,2) - \Sigma(1,2)
\end{align}
where $h$ is the Hamiltonian and $\tau$ a time on the Keldysh contour. We now consider the variation $\delta G$ of the Green's function with respect to some perturbation of the Hamiltonian $\delta h$. From Eq.~(\ref{eom}) we obtain
\begin{align}
 \int d2 [\delta G^{-1}(1,2)G(2,3)+ G^{-1}(1,2)\delta G(2,3)]=0
\end{align}
and thus
\begin{align}
\delta G(1,3)=&\int d2d4 G(1,2)\delta G^{-1}(2,4)G(4,3)\\
=&-\int d2d4 G(1,2)[\delta(2,4)\delta h(4)+\delta\Sigma(2,4)]\notag\\
&\times G(4,3) \label{deltaG}.
\end{align}
with $\delta \Sigma=\Sigma[G+\delta G]-\Sigma[G]$.
Equation (\ref{deltaG}) defines a recursive relation for $\delta G$ as we can write $\delta \Sigma(1,2)=\int d3d4 K(1,3;2,4)  \delta G(3,4)$, where $K=\delta^2\Phi/\delta G^2$ is the irreducible two particle vertex, which is obtained from $\Phi$ by cutting two lines.
We can reorganize Eq.~(\ref{deltaG}) to find
\begin{align}
\delta G(1,3)=&-\int d2d4 \delta(2,4)\delta h(4) L(1,2;3,4) \label{generic_response},
\end{align}
where $L(1,2;3,4)$ is a reducible vertex functions, which is related to the irreducible vertex $K$ via the Bethe--Salpeter equation
\begin{align}
 L&(1,2;3,4)=G(1,4)G(2,3)\notag \\
 &+\int d5d6d7d8 G(1,5)G(6,3)K(5,8;6,7)L(7,2;8,4)\label{bse}
\end{align}
shown in Fig.~\ref{fig:conserving}(c).

The two-particle function $L$ relates the change of the Green's function to a perturbation in the Hamiltonian and can be used to calculate arbitrary response functions. For instance, the current response to a vector potential reads
\begin{align}
 \delta J_\mu(1)=\int d2 \chi_{\mu\rho}(1,2)\delta A^\rho(2),
\end{align}
where Einstein sum convention is implied and
\begin{align}
 \chi&_{\mu\rho}(1,2)=\notag\\
 &-i\Bigl(\frac{p_{1,\mu}-p_{1',\mu}}{2m_\mu }\Bigr)\Bigl(\frac{p_{2,\rho}-p_{2',\mu}}{2m_\nu }\Bigr)L(1,2;1',2')|_{ \substack{1'\to 1\\ 2'\to 2}}\label{response_function}
\end{align}
Diagrammatically this amounts to connecting pairs of outer legs of the two-point function $L$ to current vertices as shown in Fig~\ref{fig:conserving}(d).

We now derive the transconductivity diagrams for the electron-exciton system. Our starting point is the functional $\Phi$ depicted in Fig.~\ref{fig:conserving}(a). The prefactors of the individual terms are determined by the number of symmetry operations of each diagram. The choice of the diagram is motivated by the ladder approximation for the T matrix discussed above. Summing over all possible ways to cut a single line in this diagram, we obtain the disorder self energies for excitons and electrons as well as the polaron self energy in the ladder approximation. The resulting Green's functions are displayed in Fig.~\ref{fig:Tmatrix}(a). By cutting all possible pairs of lines in $\Phi$, we obtain a set of diagrams that form the irreducible vertex $K$ shown in Fig.~\ref{fig:conserving}(b), which is kernel of the Bethe--Salpeter equation (\ref{bse}). Note that $K$ includes exciton-electron as well as exciton-exciton vertices. Here we have neglected any self energy or vertex corrections of the electronic Green's function due to interactions, anticipating that these do not contribute to the transconductivity to leading order in the polaron density. For the same reason, we restrict the response function to diagrams with a single polaron loop. Moreover, we have used the fact that vertex corrections due to disorder are absent for Gaussian white noise.

The solution of the Bethe--Salpeter equation (\ref{bse}) shown in Fig.~\ref{fig:conserving}(c) is the reducible two-particle vertex $L$, which is directly related to the response function by Eq.~(\ref{response_function}). As a result, we obtain two types of diagrams displayed in Fig.~\ref{fig:conserving}(d). We emphasize that these are the only diagrams contributing to the linear order in the polaron density $n_x$ within the conserving approximation, as any diagram with an additional internal exciton loop would yield a result $\propto n_x^2$.
These diagrams are closely related to the so-called Maki--Thompson and Aslamazov--Larkin diagrams that describe superconducting fluctuations (similar diagrams appear in Ref.~\onlinecite{Enss2011} that considers transport in a two-component Fermi gas). 
We point out that these diagrams recover the perturbative Coulomb drag results if expanded to lowest order in the interaction \cite{Kamenev1995,Flensberg1995}. Crucially, however, we here need to include vertex corrections for the polarons, which arise from the third diagram in Fig.~\ref{fig:conserving}(b). Moreover, we emphasize that the T matrices are to be evaluated self-consistently in order to remain within the conserving approximation. Indeed, one can readily verify that the self-consistent self energy and the vertex corrections at zero external frequency satisfy the Ward identity (cf.\ Sec.~\ref{sec:vertex_functions}).
Finally, we point out that, while there are no vertex corrections from disorder, the effect of impurities is included as a broadening of the electron and exciton Green's functions [see Fig.~\ref{fig:Tmatrix}(a)].

We now switch to Fourier space to make use of the  translational symmetry restored after disorder averaging. The transconductivity is related to the response function $\chi$ in Fig.~\ref{fig:conserving}(d) by
\begin{align}
\sigma_{\alpha\beta}(\Omega)=&\frac{e}{i \Omega} \chi_{\alpha\beta}(\Omega)\label{sigma}
\end{align}
and the response function reads
\begin{align}
  \chi_{\alpha\beta}(\Omega)=&-i \int \di p  \pi_{\alpha} (p;\Omega)G_x(p) G_x (p+\Omega)  \Gamma_\beta(p;\Omega) \label{K},
\end{align}
where $p+\Omega\equiv(\B p,\omega+\Omega)$. Here, we have defined the exciton current vertex
\begin{align}
\B \Gamma &(p;\Omega)  = \frac{\B p}{m_x}+\int\di k \di k' G_e(q) G_e(p+k'-k+\Omega)\notag\\
&\times T(p+k')T(p+k'+\Omega) G_x(k) G_x(k+\Omega)\B \Gamma (k;\Omega) .
 \label{excitonvertex1}
\end{align}
The vertex $\B \Gamma$ describes the renormalization of the exciton velocity due to many-body interactions (see Fig.~\ref{fig:conserving}(d)). 
The vertex function $\boldsymbol{\pi}$ describes the coupling of excitons to an electric current and is given by
\begin{align}
\boldsymbol{\pi}(p;\Omega) 
  =-i& \int \di k \frac{ \B k }{m_e} G_e(k) G_e(k+\Omega)\Bigl[T(p+k+\Omega) \notag\\
 &+i \int \di k' G_e(k')T(p+k')T(p+k'+\Omega) \notag\\
 &\times G_x(p+k'-k) \Bigr]\label{pi1}.
\end{align}

\subsection{Low exciton density expansion}
\label{sec:low_density}

Before embarking on the evaluation of the response function, we clarify the regime of validity of our calculation. As mentioned above we are interested in the result to linear order in the exciton density $n_x\propto \mu_x^*$. Nevertheless, we assume both electron and exciton Fermi levels to exceed the frequency and disorder scattering rate, which allows us to linearize the dispersions around the Fermi levels, and neglect localization effects. To be precise we consider the limiting case $\Omega, 1/\tau_x \ll \mu_x^*$ and $\Omega, 1/\tau_e \ll \mu_e$. At the same time we assume $\mu^*_x\ll \mu_e$, which ensures that the Fermi-polaron picture remains valid.

Within this approximation we can simply set $T(p+\Omega)\simeq T(p)$ to lowest order in $\Omega$. The external frequency thus only enters in products of Green's functions, $G_x(p)G_x(p+\Omega)$ and $G_e(k)G_e(k+\Omega)$, where $\Omega$ separates the branch cuts of the two Green's functions. We can now write the exciton current vertex as
\begin{align}
 \label{excitonvertex}
\B \Gamma (p;\Omega)  =& \frac{\B p}{m_x}+\int\di k w(p,k) G_x(k) G_x(k+\Omega)\B \Gamma (k;\Omega) ,
\end{align}
where we introduced the kernel
\begin{align}
 w(p,k)=\frac{\delta \Sigma_{\rm int}(p)}{\delta G_x(k)}=
 \int \di q G_e(q) G_e(p+q-k)  T^2(p+q)\label{W_kernel},
\end{align}
which corresponds to the last term in Fig.~\ref{fig:conserving}(b). 
The vertex function $\boldsymbol{\pi}$ reads
\begin{align}
\boldsymbol{\pi}(p;\Omega) &=-i \int \di k \frac{ \B k }{m_e} G_e(k) G_e(k+\Omega)\frac{\delta\Sigma_{\rm int}(p)}{\delta G_e(k)} \label{pi_sigma}\\
&  =-i \int \di k \frac{ \B k }{m_e} G_e(k) G_e(k+\Omega)\Bigl[T(p+k) \notag\\
 +i &\int \di k' G_e(k')T(p+k')^2 G_x(p+k'-k) \Bigr]\label{pi}
\end{align}
We can alternatively apply the vertex corrections to the vertex $\boldsymbol{\pi}$ and rewrite Eq.~(\ref{K}) as
\begin{align}
\chi_{\alpha\beta}(\Omega)=&-i \int \di p  \Pi_{\alpha} (p;\Omega)  G_x(p) G_x (p+\Omega) \frac{ p_\beta}{m_x}
\end{align}
where we introduced the dressed vertex $\B \Pi$ defined by
\begin{align}
\B \Pi(p;\Omega)  &= \boldsymbol{\pi}(p;\Omega)\notag\\
&+\int \di k \B \Pi (k;\Omega) G_x(k) G_x(k+\Omega)w(k,p).
\label{Pii}
\end{align}

We now outline our strategy for expanding Eqs.~(\ref{K}) and (\ref{excitonvertex}) to linear order in $n_x$. While we have restricted the calculation to diagrams with only a single exciton loop, thereby neglecting certain higher order contributions, single-loop diagrams may still contain terms nonlinear in $n_x$ that should be eliminated. Our starting point is the following decomposition of the exciton Green's function
\begin{align}
G_x(p) = G_x^R(p) + 2 i \mathrm{Im} G_x^A (p) \theta(-\omega).  \label{Gx_expansion}
\end{align}

In the diagrams, each Green's function $G_x$ appears inside a frequency integration and we can write
\begin{align}
 \int dp  f(p)G_x (p)=& \int dp  f(p)G_x^R (p)\notag\\
 &+\int dp f(p)2i\mathrm{Im} G_x^A (p) \theta(-\omega),\label{G_expansion}
\end{align}
where the second term is proportional to $n_x$ for any function $f(p)$ that does not have poles in the lower half-plane $\omega<0$. Hence, we can use Eq.~(\ref{G_expansion}) to expand a loop diagram in powers of $n_x$. The zeroth order is given by replacing all exciton Green's functions in a loop by retarded functions. This contribution to the transconductivity trivially vanishes.
The decomposition (\ref{G_expansion}) thus suggests a simple recipe to generate the diagrams at first order in $n_x$: consider all diagrams, where a single line exciton line is replaced by $\mathrm{Im} G_x^A (p) \theta(-\omega) $ and all others are assumed to be retarded functions.

Indeed, this recipe works for the functions $w(p,k)$ and $\boldsymbol{\pi}(p,\Omega)$ in Eqs.~(\ref{W_kernel}) and (\ref{pi}) as
all exciton Green's functions in these expressions, including the internal Green's functions in the definition of the T matrix, are of the form of Eq.~(\ref{G_expansion}). The only exception to this simple expansion rule are the expressions in Eqs.~(\ref{K}) and (\ref{excitonvertex}), where products of two exciton Green's functions within the same frequency integral appear. The expansion of such terms will be derived in Sec.~\ref{sec:calculate_conductivity} below.

To simplify bookkeeping, we consider $\boldsymbol{\pi}$ and $w$ to be functionals of $G_x$. The expansion $\boldsymbol{\pi}\simeq \boldsymbol{\pi}^{(0)}+\boldsymbol{\pi}^{(1)}$ can then be expressed to first order in $n_x$ as
\begin{align}
\boldsymbol{\pi}^{(0)}=&\boldsymbol{\pi}[G^R_x]\\
 \boldsymbol{\pi}^{(1)}=&\int \di p \frac{\delta \boldsymbol{\pi}[G^R_x]}{\delta G^R_x(p)} 2 i \mathrm{Im} G_x^A (p) \theta(-\omega)\label{pi_1}
\end{align}
where $\boldsymbol{\pi}[G^R_x]$ means that all exciton Green's functions inside the vertex have been replaced by retarded functions. Similarly, the function $w$ can be expressed as $w\simeq w^{(0)}+w^{(1)}$ with
\begin{align}
w^{(0)}=&w[G^R_x]\\
 w^{(1)}=&\int \di p \frac{\delta w[G^R_x]}{\delta G^R_x(p)} 2 i \mathrm{Im} G_x^A (p) \theta(-\omega).\label{W_1}
\end{align}

Naively, it may seem cumbersome to expand all exciton Green's functions as described above for both diagrams shown in Fig.~\ref{fig:conserving}(d), however, we can considerably simplify the solution by symmetry arguments. Importantly, both diagrams have been derived by functional derivatives of the $\Phi$ functional and, therefore, the representation of the expansion in $n_x$ in terms of functional derivatives is particularly suitable for our problem. Most terms in the expansion simply correspond to some higher order derivatives of the $\Phi$ functional or the self energy. The fact that the order of differentiation does not matter leads to important symmetry properties, for instance, $w(p,p')=\delta \Sigma(p)/\delta G_x(p')=w(p',p)$, which can be readily verified from Eq.~(\ref{W_kernel}). Moreover, this relation is immediately obvious from the diagrammatic representation of $w$, shown as the last term in Fig.~\ref{fig:conserving}(b).

We make use of symmetry properties as well as the Ward identity in the evaluation of the transconductivity below. Indeed, we find at the end of Sec.~\ref{sec:calculate_conductivity} that the various contributions from the expansion of the self energy and vertex corrections eventually combine into a simple final expression that can be readily computed.
As mentioned above, however, the product $G_x(p)G_x(p+\Omega)$ cannot be simply expanded using functional derivatives. This term accounts for the mobility of excitons and depends sensitively on the parameter $\Omega\tau_x$. A similar expression $\sim G_e(k)G_e(k+\Omega)$ occurs in the definition of $\pi(p;\Omega)$ in Eq.~(\ref{pi_sigma}) and contains information about the electron mobility.
The latter expression can be simplified by expanding the Green's functions in terms of delta functions around the quasiparticle resonance $\epsilon=\xi_\B k$. In App.~\ref{app:proof_product} we show that the approximation
\begin{align}
 \frac{\B k}{m_e} G_e(k)G_e(k+\Omega)\simeq \frac{\left(i\tau_e\Omega \frac{\B k}{m_e}  \partial_{\epsilon} +\partial_{\B k} \right) G_e( k)}{1-i\tau_e\Omega }, \label{G_product_integral}
\end{align}
is valid to leading order in $\Omega,1/\tau_e\ll \mu_e$. We cannot immediately use this relation for the exciton Green's function because its nonperturbative interaction self-energy correction precludes a description in terms of on-shell properties only. We will discuss this issue in more detail in Sec.~\ref{sec:calculate_conductivity} below.

\subsection{\texorpdfstring{ Evaluation of the vertex functions to zeroth order in $\protect n_x$}{}}
\label{sec:vertex_functions}
We begin with the evaluation of the vertex functions $\B\Gamma(p,\Omega)$, $\boldsymbol{\pi}(p,\Omega)$ and $\B \Pi(p,\Omega)$ to zeroth order in the polaron density by simply replacing all exciton Green's functions by retarded functions. Equation~(\ref{excitonvertex}) reads at zeroth order in $n_x$
\begin{align}
 \B \Gamma^{(0)} (p)  \simeq \frac{\B p}{m_x}+\int\di k w^{(0)}(p,k) \B \Gamma^{(0)} (k) G^R_x(k)^2  , \label{Gamma_vertex_equation}
\end{align}
where we have approximated $G_x^R(k) G_x^R(k+\Omega) \to G_x^R(k)^2$ to lowest order in $\Omega$.
Using the chain rule
\begin{align}
 \partial_{\bf p}\Sigma_{\rm int}^{(0)}(p)&=\int dq \frac{\delta \Sigma_{\rm int}^{(0)}(p)}{\delta G^R_x(q)}\partial_{\B q}G^R_x(q),
\end{align}
which immediately follows from the definition of $\Sigma_{\rm int}$ and $T$ in Eqs.~(\ref{selfenergy}) and (\ref{T1}), as well as $\delta \Sigma_{\rm int}^{(0)}(p)/\delta G^R_x(q)=w^{(0)}(p,q)$, we can readily verify that the solution of equation (\ref{Gamma_vertex_equation}) is given by the Ward identity
\begin{align}
\B \Gamma^{(0)}(p)&=\frac{\B p}{m_x}+\partial_{\bf p}\Sigma_{\rm int}^{(0)}(p)\\
&=[G^R_x(p)]^{-2}\partial_{\B p}G_x^R(p).\label{Ward}
\end{align}
In order to evaluate $\boldsymbol{\pi}^{(0)}(p,\Omega)$, defined by Eq.~(\ref{pi}), we employ Eq.~(\ref{G_product_integral}) and write
\begin{align}
 \boldsymbol{\pi}^{(0)}&(p;\Omega) =i \int \di k
G_e( k)\frac{i\tau_e\Omega \frac{\B k}{m_e}  \partial_{\epsilon} + \partial_{\B k}  }{1-i\tau_e\Omega }
\Bigl[T^{(0)}(p+k) \notag\\
& +i \int \di k' G_e(k')T^{(0)}(p+k')^2 G^R_x(p+k'-k) \Bigr]\label{pi0}
\end{align}
where we have performed a partial integration. Using
\begin{align}
 \partial_p T(p)&=i \int \di k T^2(p) G_e(k) \partial_{p} G_x(p-k),\label{T_deriv}
\end{align}
which follows directly from the definition of the T matrix in Eq.~(\ref{T1}), we find that the momentum derivative in Eq.~(\ref{pi0}) drops out and we obtain
\begin{align}
\boldsymbol{\pi}^{(0)}&(p;\Omega) = \frac{-i \tau_e\Omega  }{1-i\tau_e\Omega } \int \di k \frac{\B k-\B p}{m_e}
w(k,p)   \partial_{\epsilon} G^R_x(k).\label{pi_integral}
\end{align}
The dressed vertex $\B\Pi^{(0)}(p,\Omega)$ at zero polaron density can be determined by plugging this expression into Eq.~(\ref{Pii}) and approximating $G^R_x(k)G^R_x(k+\Omega)\simeq [G_x^R(k)]^2$. Using the chain rule 
\begin{align}
 \partial_\omega\Sigma_{\rm int}^{(0)}(p)=\int \di k w^{(0)}(p,k)[G^R_x(k)]^2\partial_{\epsilon}[G^R_x(k)]^{-1}
\end{align}
and an analogous identity for $ \partial_p\Sigma_{\rm int}^{(0)}(p)$ as well as  $[G^R_x]^{-1}=\omega-\omega_\B p+(i/2\tau_x) - \Sigma^{(0)}_{\rm int}(p)$ one can readily verify that Eq.~(\ref{Pii}) is satisfied by the expression
\begin{align}
\B \Pi^{(0)}(p;\Omega)=\frac{i \tau_e\Omega  }{1-i\tau_e\Omega } \Bigl(\frac{m_x}{m_e}\partial_{\B p}+\frac{\B p}{m_e}\partial_{\omega}\Bigr) \Sigma^{(0)}_{\rm int}(p).\label{Pi_gen}
\end{align}

\subsection{\texorpdfstring{Evaluation of transconductivity to first order in $\protect n_x$}{}}\label{sec:calculate_conductivity}
With the results above we are prepared to evaluate the response function $\chi_{\alpha \beta}$ in Eq.~(\ref{K}) to linear order in $n_x$. As outlined previously, we can use the representation of the exciton Green's function in Eq.~(\ref{G_expansion}). By expanding the different terms of Eq.~(\ref{K}) separately, we obtain three contributions
\begin{flalign}
\chi_{\alpha,\beta}(\Omega) =&\chi^{i}_{\alpha,\beta}(\Omega)+\chi^{ii}_{\alpha,\beta}(\Omega)+\chi^{iii}_{\alpha,\beta}(\Omega)+ {\cal O}(n_x^2)\label{chi}
\\
\chi^{i}_{\alpha,\beta}(\Omega) =&-i \int \di p \pi^{(1)}_\alpha (p;\Omega) G_x^R(p)^2 \Gamma^{(0)}_\beta(p) \label{chi1}\\
\chi^{ii}_{\alpha,\beta}(\Omega) =&-i \int \di p \di k  \Pi^{(0)}_\alpha (p;\Omega) G_x^R(p)^2W^{(1)}(p,k)\notag\\
&\times G_x^R(k)^2
 \Gamma^{(0)}_\beta(k)  \label{chi2}\\
\chi^{iii}_{\alpha,\beta}(\Omega) =& -i \int \di p  \Pi^{(0)}_\alpha (p;\Omega)   \Gamma_\beta^{(0)} (p) G_x(p) G_x(p+\Omega) \label{chi3}
\end{flalign}
To evalute the first contribution, we use Eq.~(\ref{pi_1}) as well as the Ward identity for $\B\Gamma^{(0)}$ in Eq.~(\ref{Ward}) and obtain
\begin{align}
 \chi^{i}_{\alpha,\beta}(\Omega)=&2 \int \di p \di p' \frac{\delta \pi_\alpha^{(0)}(p;\Omega)}{\delta G^R_x(p')}\mathrm{Im} G_x^A(p')\theta(-\omega')\notag\\
  &\times \partial_{ p_\beta} G_x^R(p).
\end{align}
Writing the first order term in $n_x$ as a functional derivatives turns out to be very useful. We first use Eq.~(\ref{pi_sigma}) to express the vertex $\pi_\alpha(p)$ in terms of the self energy $\Sigma_{\rm int}$ and we subsequently have to evaluate $\delta \Sigma_{\rm int}(p)/\delta G_x(p')=W(p,p')$. From its definition in Eq.~(\ref{W_kernel}), however, we immediately observe that the function $W(p,p')$ is symmetric under exchange of momentum arguments, which implies the simple relation
\begin{align}
\frac{\delta \pi_\alpha(p;\Omega)}{\delta G_x(p')}=\frac{\delta \pi_\alpha(p';\Omega)}{\delta G_x(p)}.
\end{align}
Making use of this expression together with the chain rule
\begin{align}
 \int dp'\frac{\delta \pi_\alpha(p;\Omega)}{\delta G_x(p')}\partial_{\bf p'}G_x(p')=\partial_{\bf p} \pi_\alpha(p;\Omega),
\end{align}
we find
\begin{align}
 \chi^{i}_{\alpha,\beta}(\Omega) =&2 \int \di p  \mathrm{Im} G_x^A(p)\theta(-\omega)\partial_{ p_\beta} \pi_\alpha^{(0)}(p,\Omega).
\end{align}

The structure of the second term given by Eq.~(\ref{chi2}) is similar to the first contribution. Using the Ward identity as well as the expansion of $w$ in Eq.~(\ref{W_1}) we obtain
\begin{align}
 \chi^{ii}_{\alpha,\beta}(\Omega) =&2 \int \di p \di k \di k' \Pi^{(0)}_\alpha (p;\Omega) G_x^R(p)^2\theta(-\epsilon')
 \notag\\
&\times  \mathrm{Im} G_x^A (k') 
 \frac{\delta w^{(0)}(p,k)}{\delta G^R_x(k')} \partial_{k_\beta} G_x^R(k).
\end{align}
We now use the identities
\begin{align}
  \frac{\delta w(p,k)}{\delta G_x(k')}= &\frac{\delta^2 \Sigma_{\rm int}(p)}{\delta G_x(k)\delta G_x(k')}=\frac{\delta w(p,k')}{\delta G_x(k)},\\
  (\partial_{\B p}+\partial_{\B k})w(p,k)=&\int dk' \frac{\delta w(p,k)}{\delta G_x(k')}\partial_{\B k'} G_x(k'),
\end{align}
where the second line immediately follows from the definition of $T$ and $w$  in Eqs.~(\ref{T1}) and (\ref{W_kernel}). We arrive at
\begin{align}
 \chi^{ii}_{\alpha,\beta}(\Omega) =&2 \int \di p \di k \Pi^{(0)}_\alpha (p;\Omega) G_x^R(p)^2\theta(-\epsilon)\notag\\
&\times 
   \mathrm{Im} G_x^A (k) (\partial_{p_\beta} +\partial_{k_\beta} )w^{(0)}(p,k).\label{chi2a}
\end{align}

The third contribution $\chi^{iii}$ in Eq.~(\ref{chi}) contains a term $G_x(p)G_x(p+\Omega)$ which cannot be expanded by the simple recipe in Eq.~(\ref{G_expansion}) because two exciton Green's functions are evaluated at nearby frequencies. A similar expression for the electronic Green's function has been evaluated in Eq.~(\ref{G_product_integral}), however, we cannot use the same formula for excitons because their self energy has an energy dependent imaginary part. In particular, when evaluated away from the polaron resonance at $\omega=\zeta_\B p$, ${\rm Im}\Sigma_{\rm int}(p)$ is not necessarily small. Hence, the imaginary part of the exciton Green's function cannot be simply replaced by a delta function at the polaron resonance and the energy integral has both on-shell and off-shell contributions.

The evaluation of the on-shell contribution, where $G_x\simeq\bar{G}_x$ given by Eq.~(\ref{Gatt}), is further complicated by the momentum dependent lifetime broadening $1/2\tau_x+{\rm Im}\Sigma_{\rm int}(p)$ of the polarons. For simplicity, we neglect this momentum dependence in the following, writing
\bea
\bar{G}_x (p) \simeq \frac{1}{\omega - \zeta_\B p +i / 2 \tau^*_x{\rm sgn}(\omega)} \label{reduced}
\eea 
with $1/\tau^*_x=Z/\tau_x$, which is formally justified in the limit ${\rm Im} \Sigma_{\rm int}(0,-\mu^*_x)\ll 1/{2\tau_x}$. This approximation ignores the effect of electron-polaron scattering on transport, which is expected to be suppressed at small external frequencies due to the small available phase space. We discuss this issue in more detail in Sec.~\ref{sec:low-energy} below.  

The distinction between on- and off-shell contributions can now be made explicit by writing
\begin{align}
{\rm Im}G_x(p)=Z{\rm Im}\bar{G}_x+|G^R(p)|^2{\rm Im}\Sigma_{\rm int}(p,\omega).\label{G_on_shell}
\end{align}
For the relevant energies $\omega<0$, the first term is entirely determined by on-shell contributions, whereas the second term vanishes near the polaron resonance. With the help of Eq.~(\ref{Gx_expansion}), we find
\begin{align}
 G_x(p)&G_x(p+\Omega)\simeq[G^R_x(p)+2iZ{\rm Im}\bar{G}_x(p)\theta(-\omega)]\notag\\
 &\times[G^R_x(p+\Omega)+2iZ{\rm Im}\bar{G}_x(p+\Omega)\theta(-\omega)]\notag\\
 &+4iG_x^R(p)|G^R(p)|^2{\rm Im}\Sigma_{\rm int}(p,\omega)\theta(-\omega),
\end{align}
where we have approximated $\omega+\Omega\to \omega$ in the second term and we have neglected terms of order ${\rm Im}\Sigma_{\rm int}(p)^2\propto n_x^2$. Using this expression, we can rewrite Eq.~(\ref{chi3}) as
\begin{align}
 \chi^{iii}_{\alpha,\beta}(\Omega) &= -i \int \di p  \Pi^{(0)}_\alpha (p;\Omega)   \Gamma_\beta^{(0)} (p) [Z^2\bar{G}_x(p) \bar{G}_x(p+\Omega)\notag\\
 &+4i{\rm Re}G_x(p)|G^R_x(p)|^2{\rm Im}\Sigma_{\rm int}(p)\theta(-\omega)] ,\label{chi3a}
\end{align}
where we have replaced $G_x\to Z\Bar{G}_x$ in the first term as this integral is dominated by on-shell contributions. The second term has only off-shell contributions and we can therefore write to leading order ${\rm Re}G_x(p)|G^R_x(p)|^2\simeq G^R_x(p)^3$. We obtain
\begin{align}
 \chi^{iii}_{\alpha,\beta}(\Omega) =& -i \int \di p  \Pi^{(0)}_\alpha (p;\Omega)   \Gamma_\beta^{(0)} (p) [Z^2\bar{G}_x(p) \bar{G}_x(p+\Omega)\notag\\
 &+4i[G^R_x(p)]^3{\rm Im}\Sigma_{\rm int}(p)\theta(-\omega)] .\label{chi3b}
\end{align}
To make a connection with Eq.~(\ref{chi2a}), we express ${\rm Im}\Sigma(p)$ in terms of ${\rm Im } G_x(p)$ using straightforward manipulations (see App.~\ref{app:polaron_density}), and employing the Ward identity we arrive at 
\begin{align}
 \chi^{iii}_{\alpha,\beta}(\Omega) &=  \int \di p  \Pi^{(0)}_\alpha (p;\Omega)   \{-i \Gamma_\beta^{(0)} (p)Z^2\bar{G}_x(p) \bar{G}_x(p+\Omega)\notag\\
 &+2{\rm Im}G_x(k)\theta(-\epsilon) w^{(0)}(p,k)\partial_{\B p}[G^R_x(p)]^2\},\label{chi3c}
\end{align}
Adding all three contributions, we obtain
\begin{align}
\chi_{\alpha,\beta}(\Omega) =& -i Z^2 \int \di p  \Pi^{(0)}_\alpha (p;\Omega)   \Gamma_\beta^{(0)} (p) \bar{G}_x(p) \bar{G}_x(p+\Omega) \nonumber\\
&+2 Z \int \di p {\rm Im}\bar{G}_x(p) \theta(-\omega)\partial_{\B p_\beta} \Pi_\alpha^{(0)}(p;\Omega) \notag\\
&+2 \int \di p \di k \Pi^{(0)}_\alpha (k;\Omega) \theta(-\omega)\notag\\
&\quad \times 
   \mathrm{Im} G_x^A (p) \partial_{k_\beta}[ w^{(0)}(p,k)G_x^R(k)^2].
\end{align}
We can perform a partial integration in the last term and using $\int \di k w^{(0)}(p,k)G_x^R(k)^2\simeq-Z\partial_\omega \Sigma(\omega)\simeq1-Z$, we can write the response function in the form
\begin{align}
\chi_{\alpha,\beta}(\Omega) =& -i Z^2 \int \di p  \Pi^{(0)}_\alpha (p;\Omega)   \Gamma_\beta^{(0)} (p) \bar{G}_x(p) \bar{G}_x(p+\Omega) \nonumber\\
&+ Z \int \di p \bar{G}_x(p) \partial_{\B p_\beta} \Pi_\alpha^{(0)}(p;\Omega) ,\label{chifinal}
\end{align}
where we have used Eq.~(\ref{G_expansion}) and $\int dp G_x(p)=\int dp \bar{G}_x(p)$.

We have arrived at an expression that depends exclusively on the on-shell Green's function $\bar{G}_x(p)$. Using the explicit expression in Eq.~(\ref{reduced}), we can rewrite Eq.~(\ref{chifinal}) using Eq.~(\ref{G_product_integral}) as
\begin{align}
 \chi_{\alpha,\beta}(\Omega) =-&i Z \int \di p  \bar{G}_x(p) \partial_{\B p_\beta} \Pi_\alpha^{(0)}(p;\Omega)\notag\\
 -&i Z^2 \int \di p   \Pi^{(0)}_\alpha (p;\Omega)   \Gamma_\beta^{(0)} (p)\notag\\
&\times\frac{\left(\Omega \frac{\B p_\beta}{m_x^*}\partial_\omega+\partial_{\B p_\beta}i/\tau_x^* \right) \bar{G}_x(p)}{\Omega+i/\tau_x^*}  .
\label{TransFinal}
\end{align}
Moreover, the vertex functions in Eqs.~(\ref{Ward}) and (\ref{Pi_gen}) can be evaluated explicitly 
\begin{align}
\B \Gamma^{(0)}(\B p,\zeta_\B p)&=\frac{\B p}{Z m_x^*}\label{Gamma0},\\
\B \Pi^{(0)} (\B p,\zeta_\B p;\Omega) &= \frac{-i\tau_e \Omega  }{1-i\tau_e\Omega }\frac{\B p}{Zm_e} \Bigl(1-\frac{m_x}{m_x^*}\Bigr).\label{Pi0}
\end{align}
Using these expressions and the identities $-i \int \di p \bar{G}_x(p) = n_x$ and $\int \di \omega \partial_\omega \bar{G}_x(\omega)=0$ as well as integration by parts, we readily obtain
\begin{align}
 \chi_{\alpha \beta}=&-
\frac{ \delta_{\alpha\beta}n_x}{m_e} \Bigl(1-\frac{m_x}{m_x^*}\Bigr)
\frac{\tau_e \tau^*_x\Omega^2}{(1-i\tau_e\Omega)(1-i\tau^*_x\Omega ) }.
\end{align}
The longitudinal transconductivity $\sigma_{xe} = e \chi_{\alpha \alpha}/i \Omega$ thus reads
\begin{align}
\sigma_{xe}(\Omega)=& \frac{e n_{x}}{m_e}\left(1 -\frac{m_x}{m_x^*}\right)\frac{i \Omega \tau_e \tau^*_{x} }{\left(1-i \Omega \tau_e\right)\left(1-i \Omega \tau^*_{x}\right)},\label{sigma_drag}
\end{align}
which is identical to the expression in Eq.~(\ref{transconductivity}).

\subsection{Transconductivity in terms of low-energy excitations}\label{sec:low-energy}

We can rederive the result in Eq.~(\ref{chifinal}) starting from an effective theory based on attractive polarons as the only low-energy excitations of the system. The effective low-energy Hamiltonian in terms the attractive polaron operators $a_\B p$ in the absence of electric fields reads
\bea
H_0 = \sum_\B p \zeta_\B p a^\dagger_\B p a_\B p + \sum_{\B p,\B q, j} U e^{i \B q \B r_j} a^\dagger_{\B p+\B q} a_\B p,
\eea
where the first term denotes the dispersion of polarons, while the second term corresponds to disorder scattering of polarons. The effective polaron Green's function can be written as $\bar{G}_x(p) = \int \di \omega e^{-i\omega t} \bra{0} T a_\B p(t) a^\dagger_\B p(0) \ket{0}$. 

In the presence of an electric field $\B E(t)=\B Ee^{-i\Omega t}$, the polaron quasiparticles are no longer eigenstates of the system as the field induces a drift in the Fermi sea.  Solving the polaron problem with average electron velocity $\B v_e$, we find that the polaron dispersion is shifted in momentum space $\B p\to \B p+\B v_e(m_x^*-m_x)$ (see App.~\ref{app:polaron_dispersion} for a detailed calculation). This shift can be interpreted as an effective vector potential for polarons induced by the electric field. Assuming an electronic drift velocity $\B v_e=-e\tau_e\B Ee^{-i\Omega t}/m_e(1-i\Omega\tau_e)$, we can account for the shifted dispersion by introducing an additional term in the effective polaron Hamiltonian
\bea
H'=&-\sum_\B p  a^\dagger_\B p a_\B p  Z \B \Pi^{(0)}(\B p,\zeta_\B p;\Omega)  \cdot \B A(t)\label{H1},
\eea
where $\B \Pi^{(0)}$ is the vertex evaluated at vanishing exciton density given by Eq.~(\ref{Pi0}) and $\B A(t)=\B E e^{-i\Omega t}/i \Omega$ is the electric vector potential. 

Alternatively, Eq.~(\ref{H1}) can be derived by evaluating the effective electron current vertex of polarons. Following the same arguments as in Sec.~\ref{sec:conserving}, one can readily convince oneself that this vertex is given by $Z\B \Pi = \frac{\partial}{\partial \B A} \Sigma_{\rm int}[G_e(\B A)] \Bigl|_{\B A=0}$. The evaluation of the vertex $\B \Pi^{(0)}$ is straightforward and has been performed in Sec.~\ref{sec:vertex_functions}.

The attractive polaron current resulting from the Hamiltonian $H'$ is 
\begin{align}
 \hat{\B J}_x=& \sum_\B p \left[\frac{\B p}{m_x^*}+    \frac{e \B E e^{-i\Omega t}}{i \Omega} Z  \partial_{\B p}   \Pi^{(0)}(\B p,\zeta_\B p;\Omega)\right]   a^\dagger_\B p a_\B p  \notag \\
=& \hat{\B j}_x +  \frac{e \B E e^{i\Omega t}}{i \Omega} Z    \partial_{\B p}  \Pi^{(0)}(\B p,\zeta_\B p;\Omega) \hat{n}_x \label{J_x}
\end{align}
where we introduced $\hat{\B j}_x \equiv \sum_\B p (\B p/m_x^*)  a^\dagger_\B p a_\B p $ and we used the fact that $\partial_{\B p}  \Pi^{(0)}(\B p,\zeta_\B p;\Omega)$ does not depend on momentum. The first term corresponds to the paramagnetic contribution that can be evaluated using Kubo's formula. The second term is the diamagnetic contribution that originates from the change in polaron velocity due to the shift of the dispersion implied by Eq.~(\ref{H1}). These two terms precisely recover Eq.~(\ref{chifinal}).

We can hence interpret Eq.~(\ref{chifinal}) as the para- and diamagnetic contributions to the conductivity in terms of effective polaron quasiparticles with propagator $\bar{G}_x$. In the derivation of this equation in Sec.~\ref{sec:calculate_conductivity}, we have neglected the momentum dependence of the lifetime thereby ignoring incoherent electron-polaron scattering. Here, we have not made such an assumption, which suggests that Eq.~(\ref{chifinal}) holds even for a more general momentum-dependent lifetime.

Nevertheless the result for the transconductivity remains unchanged. Obviously, only quasiparticles within a thin shell of width $\sim\Omega$ around the Fermi energy contribute to the conductivity. In close analogy to Landau Fermi liquid theory, we find that the electron scattering rate of an attractive polaron of energy $\Omega$ is proportional to $\Omega^2$ (see App.~\ref{sec:scattering}). In accordance with our expansion to lowest order in $\Omega$, the electron-scattering lifetime ${\rm Im} \Sigma_{\rm int}(p)$ of the quasiparticles relevant for transport can therefore be neglected.

\section{Transconductivity of polaron polaritons}\label{sec:polaritons}

\begin{figure}[t]
\includegraphics[width=0.99\columnwidth]{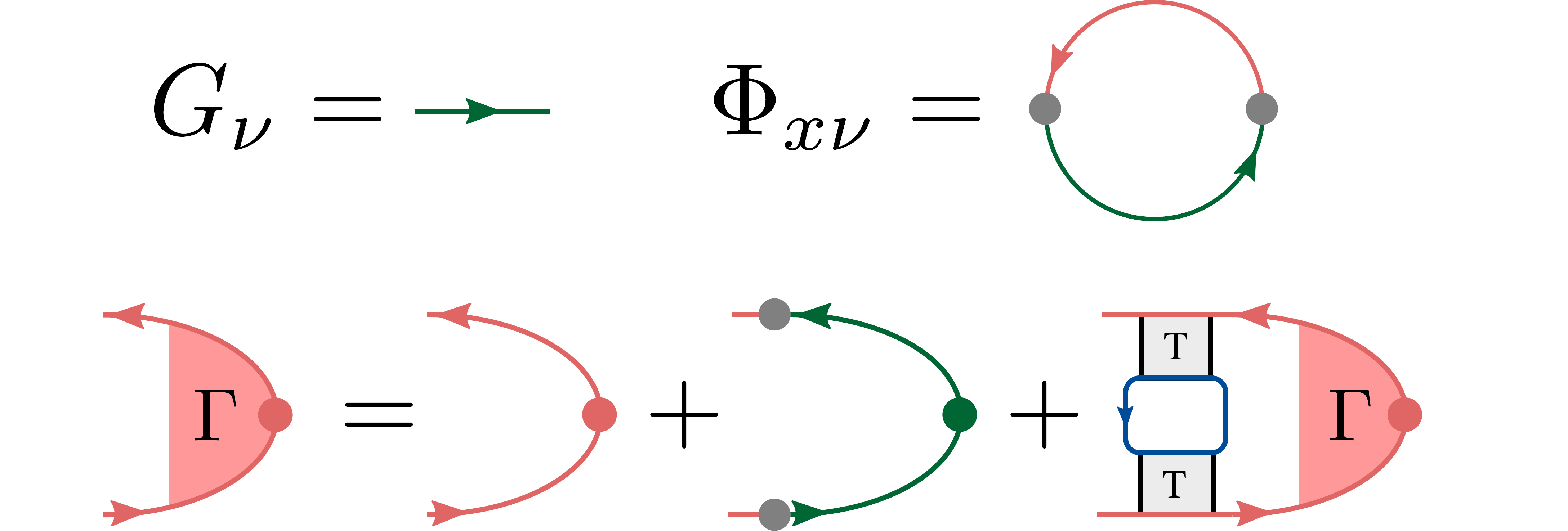}
\caption{Functional describing exciton-photon interaction $\Phi_{x\nu}[G_x,G_\nu]$ and vertex correction $\B \Gamma$ for polaritons with photon propagators represented by green (dark gray) lines.}
\label{fig:photon}
\end{figure}

The above calculation can be readily generalized to the case of exciton polaron polaritons. We simply add a term $\Phi_{x\nu}$ describing the coupling of excitons to the cavity mode to the functional $\Phi$ discussed in Sec.~\ref{sec:conserving}. The term is depicted in Fig.~\ref{fig:photon} and reads explicitly
\begin{align}
\Phi_{x\nu}=g^2 \int d1 d2 G_x(1,2) G_\nu(2,1),
\end{align}
where the photon propagator is defined as:
\begin{align}
  G_\nu(k)^{-1} = \omega-\nu_\B k +i 0\, {\rm sgn}(\omega)
\end{align}
and we introduced the dispersion of cavity photons, $\nu_\B k=\B k^2 / 2 m_\nu+\Delta$, where $m_\nu\simeq 10^{-5}m_e$ and we assume for simplicity that photons have an infinite lifetime. The functional $\Phi_{x\nu}$ leads to an additional self energy for the exciton
\begin{align}
\Sigma_{x\nu}(k)=g^2 G_\nu(k) =\frac{g^2}{ \omega-\nu_\B k +i 0\, {\rm sgn}(\omega)}\label{sigma_nu},
\end{align}
and the exciton propagator is changed accordingly
\begin{align}
G_x(p)=\frac{1}{\omega - \omega_\B p-\Sigma_{\rm int}(p)-\Sigma_{{\rm dis},x}(p) -\Sigma_{x\nu}(p)}.
\end{align}
To make connection with our previous result, we assume that we can describe the exciton as an attractive polaron neglecting other excitations such as the trion. In our approach this is formally justified if $g$ is much smaller than the energy difference between attractive polaron and trion. Nevertheless, we expect our results to be valid also at somewhat stronger couplings because, in reality, the coupling strength between cavity photons and trions is vanishingly small, even though this fact is not captured by our simple model.

Hence, approximating the bare exciton Green's function by the attractive polaron Green's function in Eq.~(\ref{Gatt}), we find
\begin{align}
G_x(p)\simeq\frac{Z}{\omega - \zeta_\B p-iZ{\rm Im}[\Sigma_{\rm int}(p)+\Sigma_{{\rm dis},x}(p)] -Z\Sigma_{x\nu}(p)}.
\end{align}
The resonances of $G_x$ are determined by the equation
\begin{align}
( \omega - \zeta_\B p)(\omega-\nu_\B k) =Zg^2.
\end{align}
Near zero momentum the lower-energy branch can be approximated by a quadratic dispersion
\begin{align}
 \gamma_\B p=\gamma_{ 0}+\frac{\B p^2}{2 m_\gamma}.
\end{align}
Hence, near the lower polariton resonance, $G_x$ takes the approximate form
\begin{align}
G_\gamma(p)&=\frac{Z_\gamma}{\omega -\gamma_{\B p}-iZ_\gamma{\rm Im}[\Sigma_{\rm int}(p)+\Sigma_{{\rm dis},x}(p)] }.\label{G_gamma}
\end{align}
When the cavity photon is tuned into resonance with the attractive polaron, $\Delta = \zeta_0$, we find
\begin{align}
 \gamma_0=& \Delta - g \sqrt{Z}\\
\frac{1}{m_\gamma}=&\frac{1}{2}\left(\frac{1}{m_x^*}+\frac{1}{m_\nu} \right)\\
Z_\gamma =& \frac{Z}{2}
\end{align}

We have arrived at an effective polariton Green's function that is formally identical to the attractive polaron Green's function in Eq.~(\ref{Gatt}), albeit with renormalized parameters. We emphasize, that the broadening ${\rm Im}[\Sigma_{\rm int}(p)+\Sigma_{{\rm dis},x}(p)]$ has to be calculated self-consistently and thus disorder scattering as well as electron scattering will be strongly suppressed due to the small polariton density of states $\sim m_\nu\ll m_x^*$.

The calculation of the transconductivity between electrons and polaron polaritons is closely related to the one for polarons.
The relation for the exciton current vertex in Eq.~(\ref{excitonvertex}) acquires an additional term and reads
\begin{align}
\B \Gamma (p;\Omega)  =& \frac{\B p}{m_x}+g^2G_\nu(p)G_\nu(p+\Omega)\frac{\B p}{m_\nu}\notag\\
&+\int\di k W(p,k) G_x(k) G_x(k+\Omega)\B \Gamma (k;\Omega).
\end{align}
This equation is also displayed in Fig.~\ref{fig:photon}. Following the same steps as in Sec.~\ref{sec:vertex_functions}, we can verify that the vertex at zeroth order in the polariton density satisfies the Ward identity
\begin{align}
\B \Gamma^{(0)}(p)&=[G^R_\gamma(p)]^{-2}\partial_{\B p}G_\gamma^R(p).
\end{align}
For small wavevectors $|\B p|\ll\sqrt{m_\gamma g}$, where $G_x(p)\simeq G_\gamma(p)$ we have
\begin{align}
 \B \Gamma^{(0)}(p)&=\frac{\B p}{Z_\gamma m_\gamma}.\label{polariton_vertex}
\end{align}
In contrast, the derivation of the vertex function $\B \Pi^{(0)}(p;\Omega)$ in Eq.~(\ref{Pi_gen}) remains unchanged. Moreover, the real part of the self energy ${\rm Re}\Sigma_{\rm int}(p)$ is largely independent of the cavity coupling and hence the on-shell expression for $\B \Pi^{(0)}(p;\Omega)$ Eq.~(\ref{Pi0}) is retained.

In Sec.~\ref{sec:calculate_conductivity}, most expressions involving $G_x$ remain unchanged as they involve an integral over a large area in momentum space. The only exceptions occur after Eq.~(\ref{G_on_shell}) where the on-shell Green's function $\bar{G}$ of occupied states is used. In these expressions we can thus simply substitute $Z\bar{G}_x$ with $Z_\gamma\bar{G}_\gamma$. These changes result in an additional factor $Z_\gamma/Z$ in the final result Eq.~(\ref{sigma_drag}). In addition, the lifetime $\tau_x$ is replaced by the polariton lifetime $\tau_\gamma=\tau_x m_x^*/m_\gamma\gg \tau_x$. Based on these considerations, we finally arrive at the transconductivity between electrons and polaritons
\begin{align}
\sigma_{\gamma e}(\Omega)=& \frac{Z_\gamma}{Z}\frac{e n_{x}}{m_e}\left(1 -\frac{m_x}{m_x^*}\right)\frac{i \Omega \tau_e \tau_{\gamma} }{\left(1-i \Omega \tau_e\right)\left(1-i \Omega \tau_{\gamma}\right)}.
\end{align}
The additional factor $Z_\gamma/Z$ takes into account the fact that the drag requires a finite excitonic quasiparticle weight. At resonance this factor reduces the polaron velocity to half its value. Moreover, we emphasize that polariton drag is much less affected by excitonic disorder because the small density of states of polaritons suppresses disorder scattering.

\section{Nonequilibrium effects}
\label{sec:nonequilibrium}

We have seen that our intuitive picture of polaron drag developed in Sec.~\ref{sec:heuristic} correctly reproduces the results of the fully microscopic model presented in Sec.~\ref{sec:diagrammatics}. This is encouraging as the semiclassical theory can be extended to include other effects that could not be captured within the linear-response calculation, but that are potentially relevant for experiments. Most notably, optically excited excitons have a finite lifetime due to recombination processes, which requires a nonequilibrium calculation.

For exciton polarons in monolayer TMDs, an ultrashort radiative lifetime of $\sim 1\,$ps for low momentum excitons implies that the assumption of an equilibrium exciton gas is not justified. Moreover, disorder scattering in state-of-the-art samples is comparable to the radiative decay rate, rendering it unlikely that a spatial displacement in exciton photoluminescence induced by an applied low-frequency electric field can be observed.

In general, we envision three different experimental scenarios, where our findings are potentially observable:

(i) In the case of interlayer excitons in TMD heterobilayers, where electrons and holes occupy conduction and valence band states in different monolayers, the exciton lifetime can be tuned electrically and can well exceed $100$~ns. Since timescales for disorder scattering are considerably shorter, we expect the interlayer excitons to be in equilibrium. Provided that the spatially indirect trion state remains bound, our results could also be used to describe drag of indirect exciton polarons, where disorder or electron scattering times can be shorter than the radiative lifetime. In this case, the resonantly generated polarons will be scattered to momentum states outside the light cone, where they can decay nonradiatively or by phonon-assisted radiative decay. The nonzero drag velocity may be detected in photoluminescence.  

(ii) Alternatively, our results may be relevant to heterobilayers at very large electron density. In this limit, screening of the interaction between valence band holes and conduction band electrons has to be taken into account, which invalidates our assumption that excitons can be regarded as rigid quasiparticles. Instead, we may consider the valence band hole as a quantum impurity interacting with a Fermi sea of conduction band electrons. For sufficiently high electron densities, it may become favorable for the hole to form a polaron rather than an exciton. This approach bears some similarity with Ref.~\onlinecite{pimenov2017fermi} and it can be used to analyze the Fermi-edge singularity problem within the framework of Fermi polarons. Our analysis of the transport problem carries over to the case of hole polarons with the trion binding energy replaced by the screened exciton binding energy. To ensure that $\Omega \tau > 1$ is satisfied, it may be possible to use microwave or Terahertz irradiation and monitor the polaron response as sidebands in optical response.

(iii) Arguably the most promising platform for observation of electric-field induced displacement of neutral optical excitations is provided by exciton polaron polaritons observed when a monolayer with a 2DES is embedded inside a 2D microcavity\cite{liu2015strong,dufferwiel2017valley,sidler2017fermi}, or when a monolayer is embedded in a dielectric structure that supports in-plane propagating photonic modes\cite{hu2017imaging}. Small-momentum excitations in the lower-energy polaron-polariton branch have two striking features: first, due to their extremely small effective mass, polaritons are to a large extent protected from disorder scattering. This effect has been observed in exciton polaritons in 1990s in GaAs heterostructures \cite{whittaker1996motional,savona1997microscopic}. Second, low-momentum polaritons can only decay radiatively through cavity mirror loss: it is therefore possible to ensure that the polariton lifetime is much longer than that of excitons by using high quality-factor cavities. Nevertheless, interactions between polaritons are also weak and we cannot expect a low density of polaritons to thermalize, rendering it essential to develop a nonequilibrium description of transport.

\subsection{Boltzmann equation}

Our aim is to develop a kinetic theory for the distribution function $g_\B k(\B r, t)$ of exciton polarons, including the effects of pumping, recombination, disorder as well as a nonzero electron drift velocity $\B v_e(t)$ due to an electric field $\B E(t)$. In the most general case, we can write the Boltzmann equation as
\begin{align}
 \frac{\di g_\B k(\B r,t) }{\di t}&=\frac{\partial g_\B k(\B r,t)}{\partial t}+\dot{\B k}\frac{\partial g_\B k(\B r,t)}{\partial \B k}+\dot{\B r}_\B k\frac{\partial g_\B k(\B r,t)}{\partial \B r}\label{Boltz1}
\end{align}
Importantly, the electric field does not exert a direct force on polarons, whose canonical momentum $\B k$ is conserved. Instead, it shifts the polaron dispersion by changing the electron velocity. A straightforward solution of the polaron problem discussed in Sec.~\ref{sec:polaron_problem} in the presence of an electron drift (see App.~\ref{app:polaron_dispersion}) yields the dispersion
\begin{align}
 \tilde{\zeta}_\B k(t) = \zeta_{\B k+(m_x^*-m_x) \B v_e(t)},
\end{align}
which is shifted from the equilibrium dispersion $\zeta_{\B k}=\B k^2/2m_x^*$ such that the polaron state at $\B k=0$ has a velocity $(1-m_x/m_x^*) \B v_e(t)$. Even though an electric field does not affect the conjugate momentum of the polaron, $\dot {\B k}=0$, it changes its kinetic momentum $m_x^* \B v_x(\B k,t)$ with the polaron velocity
\begin{align}
\B v_x (\B k,t) = \frac{\partial \tilde{\zeta}_\B k(t)}{\partial \B k} = \frac{\B k}{m_x^*}+\left(1-\frac{m_x}{m_x^*}\right) \B v_e(t). \label{vx}
\end{align}
For simplicity, we assume a spatially homogeneous distribution, $\partial_{\B r}g_\B k(\B r,t)=0$ and we suppress the spatial dependence in the following. Hence, the distribution function does not have an implicit time-dependence and Eq.~(\ref{Boltz1}) can be written as
\begin{align}
\frac{\di g_\B k(t) }{\di t}&=R \delta (\B k)-\frac{g_\B k(t)}{\tau_r}+ \left(\frac{\partial g_\B k(t)}{\partial t}\right)_{{\rm dis}}+\left(\frac{\partial g_\B k(t)}{\partial t}\right)_{{\rm int}}. \label{Boltz}
\end{align}
The first term is due to  pumping of polarons at a rate $R$ in the polaron state at $\B k=0$ by resonant laser absorption. The incidence angle of a collimated laser field determines the in-plane momentum of the exciton polarons, which is in turn much smaller than the other characteristic momentum scales in the problem such as $k_F$ and $m_e \B v_e$. By tuning the frequency of a normal-incidence single-mode laser field, it is possible to ensure that only $\B k=0$ attractive polarons can be created. The second term in Eq.~(\ref{Boltz}) corresponds to the loss of attractive polarons due to the recombination processes at a rate $1/\tau_r$. In general $\tau_r$ is expected to depend on momentum, since the recombination rate should be strongest for small momenta $\B k$ that lie inside the light-cone, whereas the decay from states outside the light cone requires the generation of additional excitations such as phonons. Here we neglect the momentum dependence of $\tau_r$, for simplicity, although the generalization of our results to include this effect is straightforward. The last two terms in Eq.~(\ref{Boltz}) conserve the number of polarons and correspond to collision processes, either due to exciton disorder or incoherent scattering off electrons. These terms will be discussed in more detail below. 

Integrating Eq.~(\ref{Boltz}) over momentum space, we obtain the time-evolution of the exciton density $n(t) \equiv n_x(t)=\int (\di \B k/4\pi^2) g_\B k (t)  $ as
\bea
\dot n(t) = R - \frac{n(t)}{\tau_r} \label{density},
\eea
where we have used that the collision integrals conserve the number of polarons.
We are moreover interested in evaluating the exciton current density defined as
\bea
n(t)\bar{\B v}_x(t) =\int \frac{\di \B k}{(2\pi)^2} g_\B k(t) \B v_x(\B k,t)  .  \label{Ptot}
\eea
Differentiating with respect to time, we obtain
\begin{align}
n(t)  \frac{\di \bar{\B v}_x(t)}{\di t}+ \dot{n}(t) \bar{\B v}_x(t) =& \int \frac{\di \B k}{(2\pi)^2} \Bigl[g_\B k(t) \frac{\di \B v_x(\B k,t)}{\di t}\notag\\
&  +\frac{\di g_\B k(t)}{\di t} \B v_x(\B k,t)\Bigr] \label{Ptotevolution}
\end{align}
The first term on the right-hand side can be readily evaluated using Eq.~(\ref{vx}) and $\dot{\B k}=0$ as
\begin{align}
\int \frac{\di \B k}{(2\pi)^2} g_\B k(t) \frac{\di \B v_x(\B k,t)}{\di t}
&=n(t) \frac{m_x^*-m_x}{m_x^*m_e} \B F_e(t)\label{RHS2}
\end{align}
with $\B F_e=m_e \dot{\B v}_e(t)$. 
The second integral in Eq.~(\ref{Ptotevolution}) can be evaluated using the Boltzmann equation~(\ref{Boltz}) and, with the help of Eq.~(\ref{density}), we obtain
\begin{align}
m_x^* &\frac{\di \bar{\B v}_x(t)}{\di t}=\frac{R}{n(t)}[ \left(m_x^*-m_x\right) \B v_e(t)- m_x^*\bar{\B v}_x(t)] \notag\\
&+ \frac{m_x^*-m_x}{m_e} \B F_e(t)+\B F_{\rm dis}(t)+\B F_{\rm int}(t)\label{total},
\end{align}
In this expression, the first and second term account for polarons in the $\B k=0$ state with velocity $\B v_x (0,t) = (1-m_x/m_x^*)\B v_e(t)$ that are generated by the laser or lost by the recombination of excitons, respectively. The third term corresponds to the drag force acting on the exciton system due to the polaronic coupling to electrons, which was the main focus of the equilibrium calculation in Sec.~(\ref{sec:diagrammatics}). 
The last two terms correspond to the friction due to exciton disorder and incoherent scattering with electrons
\begin{align}
 \B F_{\rm dis}(t) =&\frac{m_x^*}{n(t)} \int \frac{\di \B k}{(2\pi)^2} \left(\frac{\partial g_\B k}{\partial t}\right)_{\rm dis} \B v_x(\B k,t)\label{Fdis},\\
\B F_{\rm int}(t) = &\frac{m_x^*}{n(t)} \int \frac{\di \B k}{(2\pi)^2} \left(\frac{\partial g_\B k}{\partial t}\right)_{\rm int} \B v_x(\B k,t)\label{drag}.
\end{align}
These forces depend sensitively on the polaron distribution and can lead to nonlinear effects in the time evolution. We discuss them in more detail in the next section.

\subsection{Estimation of the friction forces from disorder and incoherent scattering with electrons}\label{sec:Fdis}
Friction from excitonic disorder can be described by the collision integral
\begin{align}
  \left(\frac{\partial g_\B k}{\partial t}\right)_{\rm dis}=\int \frac{\di \B k'}{(2\pi)^2}  [g_{\B k'}(t)-g_{\B k}(t)]  \tilde{M}_{\B k \B k'}\label{eq89}
\end{align}
where $\tilde{M}_{\B k \B k'}$ denotes the matrix element corresponding to the scattering of a polaron from state $\B k$ to state $\B k'$. In the simplest case, we can assume Gaussian white noise as in Sec.~\ref{sec:polaron_problem} and we obtain
\begin{align}
 \tilde{M}_{\B k \B k'}=\frac{1}{ \nu_x^*\tau_x^*}\delta[\tilde{\zeta}_{\B k}(t)-\tilde\zeta_{\B k'}(t)]
\end{align}
Note that the matrix elements depend on the shifted dispersion in the presence of the electric field. With this approximation, the collision integral simplifies to
\begin{align}
  \left(\frac{\partial g_\B k}{\partial t}\right)_{\rm dis}=\frac{1}{ \nu_x^*\tau_x^*}\int \frac{\di \B k'}{(2\pi)^2}  g_{\B k'}(t) \delta[\tilde{\zeta}_{\B k}(t)-\tilde\zeta_{\B k'}(t)]
  -\frac{g_{\B k}(t)}{\tau_x^*}\label{collision_dis}
\end{align}
and the friction force reads
\begin{align}
 \B F_{\rm dis}(t) =& -\frac{m_x^*\bar{\B v}_x(t)}{ \tau_x^*}.
\end{align}
Hence, for Gaussian white noise correlated disorder, the polaron disorder scattering time is also the relaxation time for the drift velocity of the exciton system. We emphasize that this result holds, even though we cannot treat the collision integral in Eq.~(\ref{eq89}) in the relaxation time approximation.

The force introduced in Eq.~(\ref{drag}) corresponds to an additional drag force originating from the residual interaction between electrons and polarons. While coherent scattering events of electrons and excitons result in polaron formation and the polaron drag phenomenon described in Sec.~\ref{sec:heuristic} and \ref{sec:diagrammatics}, incoherent collisions lead to a lifetime broadening of the polarons. This broadening appears as a term ${\rm Im}\Sigma_{\rm int}(p)$ in the Green's function description of polarons discussed in Sec.~\ref{sec:polaron_problem}.

The qualitative effect of incoherent broadening can be estimated from a simple argument by temporarily disregarding polaronic disorder.
For concreteness, we imagine a narrow distribution of polarons around zero momentum, which corresponds to the distribution shortly after the laser has been switched on. In the comoving frame of the electrons, the zero momentum polarons have a velocity $-\B v_em_x/m_x^*$ and thus are in an excited state. Excited polaron states, however, have a finite lifetime due to the interaction with electrons and will decay into lower-energy states, which also have a smaller absolute velocity. Hence, incoherent electron-polaron scattering will lead to a relaxation of the polaron velocity to zero {\em in the comoving frame}. In the lab frame, this corresponds to an acceleration of polarons until they reach a velocity $\B v_e$. This justifies the following ansatz for the electron friction force on the polarons
\begin{align}
\B  F_{\rm int}=-m_x^*\frac{\bar{\B v}_x(t)-\B v_e}{\tau_{\rm int}(t)}\label{fint}
\end{align}
with an effective time scale $\tau_{\rm int}$ that depends on details of the polaron distribution at time $t$. A rough estimate for  $\tau_{\rm int}$ is given by the interaction lifetime of an excited polaron state with velocity $-m_x\B v_e/m_x^*$. After this time, all the polarons at zero momentum have scattered at least once. In the comoving frame of electrons, the scattering probability does not have a very strong dependence on the direction of the final polaron momentum. Hence the polaron reach an average velocity $\simeq \B v_e$ already after a few scattering events, even though the time for each individual polaron to reach that velocity is expected to be much longer.

The expression for the friction force in Eq.~(\ref{fint}) captures the conventional Coulomb drag effect, e.g., in electron bilayer systems \cite{Zheng1993,Kamenev1995,Flensberg1995,narozhny2016coulomb}. Indeed, for perturbative interactions, $\B F_{\rm int}$ is the only contribution to drag. As we have shown in Sec.~\ref{sec:diagrammatics}, nonperturbative interactions result in an additional polaron drag effect that dominates at low temperatures and frequencies, whereas the drag force in Eq.~(\ref{fint}) is a subleading correction that we have neglected in the linear response calculation in Sec.~\ref{sec:diagrammatics}. We emphasize that in nonequilibrium systems or at finite temperatures, $\B F_{\rm int}$ cannot necessarily be ignored.

An explicit expression for the collision integral reads
\begin{align}
\left(\frac{\partial g_\B k}{\partial t}\right)_{\rm int}=\int \frac{\di \B k'}{(2\pi)^2}  \Bigl[&g_{\B k'} (1-g_\B k) \tilde{Q}_{\B k' \B k}\notag\\
&-g_\B k (1-g_{\B k'}) \tilde{Q}_{\B k \B k'}\Bigr] \label{eq90}
\end{align}
where the transition probability $\tilde{Q}_{\B k \B k'}$ denotes the scattering rate of an attractive polaron of momentum $\B k$ to a momentum $\B k'$ due to interactions with an electron Fermi sea drifting at velocity $\B v_e$. In order to derive an expression for $\tilde{Q}_{\B p\B p'}$, we first consider the polaron scattering rate $Q_{\B p\B p'}$ when the electron drift velocity is zero. The electron-polaron scattering amplitude is simply given by the T matrix, and, hence, Fermi's Golden rule yields
\begin{align}
 Q_{\B p\B p'} =& Z^2\pi\int \frac{\di \B k}{(2\pi)^2} |T(\B p+\B k,\zeta_\B p+\epsilon_\B k)|^2
  f_{\B k}\left[1 - f_{\B k+\B p-\B p'}\right]\notag\\
&\times \delta(\zeta_\B p-\zeta_{\B p'}-\varepsilon_{\B k+\B p-\B p'}+\varepsilon_\B k)\label{Q0}
\end{align}
where $f$ denotes the equilibrium electron distribution. One can readily verify that this scattering rate reproduces the lifetime broadening of the on-shell polaron Green's function (see App.~\ref{sec:polaron_lifetime_self_energy} for the case of an empty polaron band)
\begin{align}
{\rm Im}\Sigma_{\rm int}(\B p,\zeta_\B p)= \int \frac{\di \B k'}{(2\pi)^2}  (1-g_{\B k'}) \tilde{Q}_{\B k \B k'}.
\end{align}
In the presence of a nonzero electron drift velocity, Eq.~(\ref{Q0}) has to be modified, by shifting both the polaron dispersion $\zeta_\B p\to \tilde{\zeta}_\B p$ and the electron distribution $f_\B k\to f_{\B k-m_e \B v_e(t)}$. A straightforward calculation along the lines of App.~\ref{app:polaron_dispersion} establishes a relation between the scattering amplitude with and without and electric field
\begin{align}
\tilde{Q}_{\B p\B p'} &=Q_{\B p-m_x \B v_e(t),\B p'-m_x \B v_e(t)}.
\end{align}
Indeed, this result confirms the intuition that the friction between polarons and electrons results in the relaxation of the polaron velocity in the comoving frame of electrons. To see this, we observe that the lowest energy state has an infinite lifetime, i.e., $Q_{0,\B k}=0$. In the presence of an electric field, this means a polaron state with momentum $\B p=m_x\B v_e$ is stable with respect to scattering off electrons. According to Eq.~(\ref{vx}) this state has a velocity $\B v_e$, i.e., it is moving at the same speed as the electron Fermi sea.

In order to obtain an estimate for the friction force $\B F_{\rm int}$ in Eq.~(\ref{drag}) on a narrow distribution centered at zero momentum, $g_\B k\simeq n\delta(k)$, we can make the same approximation as above that the average velocity after a single scattering event is $\B v_e$. We can hence write the force from first term in Eq.~(\ref{eq90}) as
\begin{align}
\frac{m_x^*}{n(t)} &\int \frac{\di \B k}{(2\pi)^2}\frac{\di \B k'}{(2\pi)^2}  \B v_x(\B k,t)
 g_{\B k'} (1-g_\B k) \tilde{Q}_{\B k' \B k}
\simeq\frac{m_x^*\B v_e}{\tau_{\rm int}(t)}.
\end{align}
and for the second term we obtain
\begin{align}
\frac{m_x^*}{n(t)} &\int \frac{\di \B k}{(2\pi)^2}\frac{\di \B k'}{(2\pi)^2}  \B v_x(\B k,t)
 g_{\B k} (1-g_\B k') \tilde{Q}_{\B k \B k'}
\simeq\frac{m_x^*\bar{\B v}_x}{\tau_{\rm int}(t)}.
\end{align}
Here we have defined the scattering time
\begin{align}
 \frac{1}{\tau_{\rm int}(t)}=\sum_{\B k'} \tilde{Q}_{ 0 \B k'},
\end{align}
which is identical to the interaction lifetime of a polaron at momentum $-m_x\B v_e$. With these results we indeed recover Eq.~(\ref{fint}).

To obtain a rough estimate of $\tau_{\rm int}$, we can approximate the T matrix by a constant $ZT(\B p+\B k, \zeta_\B p+\varepsilon_\B k) \simeq U\equiv ZT(k_F, 0)$ which is valid in the relevant limit $|\B p|\ll k_F$ and $|\B k|\simeq k_F$. 
The scattering time at zero temperature can then be obtain from a straightforward calculation (assuming $m\equiv m_x\simeq m_e$, see App.~\ref{sec:scattering}) as
\begin{align}
  \frac{1}{\tau_{\rm int}(t)} \sim \rho_e U^2 \frac{(m v_e )^3}{k_F} \label{lifetime1}.
\end{align}
At frequencies close to the trion energy the T-matrix is dominated by the trion pole. The interaction between attractive polarons and electrons originates from virtual scattering events into a higher energy state comprising a trion and a hole which results in an effective attractive interaction. This yields the estimate
\begin{align}
U \simeq - \frac{Z \epsilon_T}{\rho_e \Delta},
\end{align}
where $\Delta$ is the energy separation between the trion-hole continuum and the attractive polaron. Assuming a quasiparticle weight $Z\simeq \sqrt{\mu_e/\epsilon_T}$, the friction force on polarons at zero momentum is of the order of 
\begin{align}
\B  F_{\rm int}\simeq \frac{  m^2k_F\epsilon_T}{\Delta^2} v_e^3\B v_e.
\end{align}
The friction force therefore scales with the electric field as $F_{\rm int}\propto E^4$, which illustrates why this effect has not been captured in the linear response calculation in Sec.~\ref{sec:diagrammatics}.

We emphasize that this result is valid at zero temperature. At finite temperatures, the friction force is still expected to have the form in Eq.~(\ref{fint}), but the scattering rate acquires an additional temperature-dependent contribution due to the enhanced phase space available for electron-polaron scattering.

\subsection{Equations of motion for polarons and polaritons}

After the pump has been switched on for a time $\sim \tau_r$, the polaron density saturates to the value $n=R\tau_r$. Even though the density has reached a steady-state value, the solution of the nonlinear Boltzmann equation (\ref{Boltz}) depends sensitively on the ratio of the various time scales and in general requires numerical calculations. To ensure a strong hybridization of polarons and photons, it is desirable to work in a limit where the radiative lifetime is the shortest time scale, $\tau_r\ll \tau_x^*,\tau_{\rm int}$. In this case, the polarons remain mostly near $\B k=0$, and therefore within the light cone, as disorder and electron scattering, which change the polaron momentum, are suppressed.

This limit guarantees that the friction caused by electrons can be described by Eq.~(\ref{fint}) and we can make the connection to the semiclassical equations in Sec.~\ref{sec:heuristic} more explicit. Assuming a density $n(t)=R\tau_r$, Eq.~(\ref{total}) can be written as
\begin{align}
m_x^* \frac{\di \bar{\B v}_x(t)}{\di t}=&\frac{\left(m_x^*-m_x\right) \B v_e(t)}{\tau_r} - \frac{m_x^*\bar{\B v}_x(t)}{\tau_r}  -\frac{m_x^*\bar{\B v}_x(t)}{ \tau_x^*}\notag\\
&+ \frac{m_x^*-m_x}{m_e} \B F_e(t)-\frac{m_x^*(\bar{\B v}_x(t)-\B v_e)}{\tau_{\rm int}(t)}. \label{Boltzmann}
\end{align}
A comparison with the corresponding Eq.~(\ref{eom_x}) in Sec.~\ref{sec:heuristic} reveals three extra contributions in the non-equilibrium case. The first term originates from pumping polarons into the state $\B k=0$ with velocity $\B v_x (\B k= 0,t) = (1-m_x/m_x^*)\B v_e(t)$. The second term describes recombination of polarons. Finally, the last term originates from incoherent scattering of electrons, which has been neglected in Sec.~\ref{sec:heuristic}. This assumption is justified in the limit of weak electric fields.

Eq.~(\ref{Boltzmann}) can be readily generalized to polaritons. As has been argued in Sec.~\ref{sec:polaritons}, polaritons are much less affected by scattering processes because of their ultra-low mass and the correspondingly small scattering phase space. Resonant coupling between polarons and the cavity mode creates a local minimum near zero momentum in the polariton dispersion. For a sufficiently strong coupling, ($g>(m_xv_e)^2/2m_x^*$ in the notation of Sec.~\ref{sec:polaritons}), disorder or interaction scattering of low-energy polaritons to large momenta $\sim m_x|\B v_e|$ are energetically forbidden. In this case the polariton scattering rates $1/\tau_x^*$ and $1/\tau_{\rm int}$ are reduced by a factor $(m_\gamma/m_x^*)$ compared to exciton polarons. As typical values are $(m_\gamma/m_x^*)\sim 10^{-5}$, we conclude that such scattering processes can be neglected. Following similar steps as above, we arrive at the equation of motion for polaritons
\begin{align}
\frac{\di \bar{\B v}_x(t)}{\di t} =&\Bigl(1-\frac{m_x}{m_x^*}\Bigr)\frac{\B v_e (t)}{2 \tau_r} - \frac{ \bar{\B v}_x(t)}{\tau_r}
 + \Bigl(1-\frac{m_x}{m_x^*}\Bigr) \frac{\B F_e(t) }{2 m_e}
\end{align}
The additional factor of $1/2$ in the first and third term reflects the ratio of the polariton and polaron quasiparticle weights. This equation can be readily solved. Assuming a static electric field, such that $\B F_e(t)=0$, the polaritons move at a velocity $\bar{v}_x =(1-m_x/m_x^*)\B v_e(t)/2$ during their entire lifetime, which corresponds to an approximate distance $(1-m_x/m_x^*)\B v_e(t)\tau_r/2$. 

\section{Magnetic field response of exciton polarons}\label{sec:magneticresponse}

We can include the effect of a dc magnetic field by adding a Lorentz force to Eq.~(\ref{eom_el}).
\begin{align}
\frac{\di }{\di t } \B v_e(t) &= \B a_e(t)= -\frac{e \B E(t)}{m_e}- \B v_e(t)\times \frac{e\B B}{m_e} -\frac{ \B v_e(t)}{\tau_e}
\end{align}
We can readily solve this and find
\begin{align}
\B v_e(\Omega) &= \frac{e }{m_e}\frac{\B E(\Omega)\times e\B B/m_e-\B E(\Omega)(\tau_e^{-1}-i \Omega)}{(\tau_e^{-1}-i \Omega)^2+\omega_c^2}
\end{align}
where $\omega_c=eB/m_e$ is the cyclotron frequency. As the magnetic field does not directly couple to excitons, Eq.~(\ref{v_x}) remains valid. Hence, when $\omega_c\gg\Omega,1/\tau_e$ and $\Omega\tau_x\gg 1$, the electrons and polarons drift in the direction perpendicular to the electric and magnetic field realizing a Hall effect of neutral excitons!

More generally, Eq.~(\ref{v_x}) predicts that excitons will follow the trajectory of electrons (scaled by a factor) on time scales shorter than the exciton impurity scattering time. In the absence of an electric field, excitons should therefore move in cyclotron orbits, which suggests that polarons could experience a phenomenon similar to Landau quantization. Equivalently, one can argue that excitons should be affected by the quantizing magnetic field as they are dressed by particle-hole excitations with a discrete energy spectrum due to electronic Landau levels.
A signature of this would be a polaron spectral function with a series of peaks on top of an incoherent background present at higher energies roughly spaced by the cyclotron frequency. A related phenomenon has been discussed in the context of Bose polarons, in particular,  for electrons strongly coupled to
dispersionless phonons. In this case, phonon shake-off processes
lead to a series of broad peaks in the polaron spectral function
separated in energy by multiples of the phonon frequency \cite{Alexandrov2010}.

Observation of Landau levels in absorption spectrum where the energy separation is given by the electron cyclotron frequency would unequivocally demonstrate the central role played by polaron physics in optical excitations out of a 2DES. If bound trions were observable in absorption, the observed Landau level spacing would be determined by the trion mass, which is a factor of 3 larger than that of the electron in TMD monolayers. In the opposite limit of very high electron densities, we expect screening to lead to ionization of excitons: in this regime, the level separation will be determined by the reduced mass of electron-hole pairs.

Conversely, the motion of electrons in a magnetic field can be
influenced by the presence of excitons. Excitons at an  appreciable
density can lead to a polaronic dressing of electrons thereby
increasing their effective mass, which could be detected in
Shubnikov--de Haas oscillations or cyclotron resonance measurements.
With increasing density of excitons, the electronic resonance
frequency is expected to shift as a result of a polaronic dressing.
In contrast, bound trions would appear as a new resonance in
addition to the bare electron resonance and no shift is expected as
a function of trion density. Such experiments could therefore
provide unique evidence for polaron formation in TMDs and clarify
the nature of the exciton ground state.

\section{Outlook and conclusion}
Our work opens up new frontiers in nonequilibrium many-body physics by showing that external electric and magnetic fields could be used to control and manipulate elementary optical excitations such as excitons or polaritons. The requisite element leading to this intriguing functionality is the presence of nonperturbative interactions between excitons and electrons, leading to the formation of exciton polarons. In addition to the potential applications in realizing effective gauge fields for photonic excitations that we already highlighted, we envision several extensions of our work that by themselves constitute open theory problems.

Arguably the most interesting extension of our work is the investigation of the degenerate Bose-Fermi mixture regime which can be accessed by increasing the optical pump strength. In the limit of perturbative electron-exciton interactions, this problem could be formulated as a coupling between Bogoliubov excitations out of a polariton condensate and electrons close to the Fermi surface\cite{cotlect2016superconductivity}. In the opposite limit of nonperturbative exciton-electron interactions and polariton density $n_{p}$ lower than that of electrons $n_e$, the ground state could be described as a polaron-polariton condensate. We expect the confluence of these two approaches to give rise to new physics where the nature of polariton screening by electrons could be dramatically modified due to degeneracy favoring a high quasi-particle weight. Moreover, it is precisely in this regime that the modification of the electronic transport properties due to degenerate-polaron formation would become significant. It is possible that the previous proposals for polariton mediated superconductivity may have to be revisited in light of new features that emerge from a rigirous analysis of polaron-polariton condensation \cite{laussy2010exciton,cotlect2016superconductivity,kavokin2016exciton}. Last but not least, a gradual increase (decrease) of polariton (electron) density from $n_p \ll n_e$  to $n_p \gg n_e$ regime takes us from a Fermi-polaron problem to a Bose-polaron problem: exciton-polariton system allows for such tuning by changing the optical pump strength together with applied gate voltages that control the electron density.

Another exciting extension of our work is the analysis of the regime of strong ac-drive of the interacting electron-polariton system where electrons occupy Floquet bands. We expect a particularly strong modification of polaron formation if the external field resonantly drives plasmon resonance of degenerate electrons. Introduction of spatial modulation of the electron density using Moire patters or surface acoustic waves could be used to engineer non-trivial band structure for electrons: it is thereby possible to realize either stationary or Floquet topological bands for electrons which will in turn modify exciton-polaron transport. An alternative strategy to investigate the interplay between topological order and polaron formation is to study optical excitations from fractional quantum Hall states.

As we highlighted earlier, application of our formalism to 2D materials is particularly exciting: on the one hand, these materials exhibit a valley pseudo-spin degree of freedom and a nontrivial band geometry with a sizable Berry curvature. On the other hand, they allow for creating hybrid materials combining different functionalities; an exciting recent example is a heterostructure based on exchange coupled semiconducting and a ferromagnetic monolayers. Optical excitations in such a system will be Bose polarons where the magnetic moment of valley excitons are screened by magnon excitations out of the ferromagnet.

\section{Acknowledgments}

We thank Wilhelm Zwerger for inspiring discussions and important input in the initial stages of this work. We further acknowledge stimulating discussions with M.\ Fleischhauer, B.\ I.\ Halperin, L.\ Jauregui, A.\ Kamenev, P.\ Kim, A.\ Levchenko, M.\ Lukin, Y.\ Oreg, H.\ Park, and A.\ Rosch. A.\ I.\ and O.\ C.\ were supported by a European Research Council (ERC) Advanced Investigator Grant (POLTDES). F.\ P.\ acknowledges financial support by the STC Center for Integrated Quantum Materials, NSF Grant No.\ DMR-1231319. E.\ D.\ was supported by the Harvard-MIT CUA, NSF Grant No.\ DMR-1308435 and by the AFOSR-MURI: Photonic Quantum Matter, award FA95501610323.


\appendix

\section{Effective attractive polaron propagator}\label{app:effective}
In this section we wish to show how, starting from the full Green's function $G_x$ one can introduce an effective (or projected) propagator, that describes the propagation of the low energy excitations (i.e. the attractive polarons). Starting from the general Green's function
\bea
G_x(p) =\frac{1}{\omega -\omega_\B p -\Sigma_{\rm int}(p) + i /2 \tau_x \sgn{\omega}}\label{A1}
\eea
we introduce the dispersion of attractive polarons as the lowest energy pole of the above, i.e.,
\bea
\zeta_\B p \equiv \omega_\B p + {\rm Re} \Sigma_{\rm int}(\B p,\zeta_\B p) \simeq \frac{\B p^2}{2 m_x^*} -\mu_x^*, \label{dispPolaron}
\eea
where we introduced the polaron mass $m_x^*$ and also introduced a new chemical potential $\mu_x^*$, measured from the bottom of the attractive polaron dispersion. The above equation is correct for small momenta $\B p\ll k_F$, while for larger momenta it will start to deviate from a quadratic dispersion. 

Expanding the self energy $\Sigma(\B p,\omega)$ to linear order in $\omega$ we can write: 
\begin{widetext}
\bea
G_x(p)&\simeq& \frac{1}{\omega -\omega_\B p -{\rm Re}\Sigma_{\rm int}(\B p,\zeta_\B p)-i {\rm Im}\Sigma_{\rm int}(\B p,\zeta_\B p)+(\omega-\zeta_\B p) \left. \partial_\omega \Sigma_{\rm int}(\B p,\omega)\right|_{\omega=\zeta_\B p} + i /2 \tau_x \sgn(\omega)}\\
&=&\frac{1}{\omega -\zeta_\B p+(\omega-\zeta_\B p) \left. \partial_\omega \Sigma_{\rm int}(\B p,\omega)\right|_{\omega=\zeta_\B p} + i /2 \tau_x \sgn(\omega) -i {\rm Im}\Sigma_{\rm int}(\B p,\zeta_\B p)}\nonumber\\
&=&\frac{Z_\B p}{\omega-\zeta_\B p+i Z_\B p/2 \tau_x \sgn(\omega) - i Z_\B p {\rm Im}\Sigma_{\rm int}(\B p,\zeta_\B p)} \nonumber
\eea
where we introduced the renormalization factor $Z_\B p^{-1} \equiv 1 - \left. \partial_\omega \Sigma_{\rm int}(\B p,\omega)\right|_{\omega=\zeta_\B p} $. It clear from this definition that in general the renormalization factor $Z_\B p $ is complex ($|Z_\B p|$ denotes the quasiparticle weight) since the self energy is complex. However as long as
\bea
 \left. \partial_\omega {\rm Im} \Sigma_{\rm int}(\B p,\omega)\right|_{\omega=\zeta_\B p} \ll {\rm Im}\Sigma_{\rm int}(\B p,\zeta_\B p)
\eea
we can neglect the complex part of $Z_\B p$. We prove that the above condition is satisfied for small momenta $|\B p|\ll k_F$ in App.~(\ref{sec:scattering}) where we explicitly evaluate the imaginary part of the self energy. In this limit we can approximate the quasiparticle weight by a constant:
\bea
Z_\B p^{-1}\simeq 1 - \left. \partial_\omega{\rm Re} \Sigma_{\rm int}(\B p,\omega)\right|_{\omega=\zeta_\B p} \simeq 1 - \left. \partial_\omega{\rm Re} \Sigma_{\rm int}(\B p_F,\omega)\right|_{\omega=0} \equiv Z^{-1} \label{ZZ}
\eea
We can also introduce the lifetime of the attractive polaron as: 
\bea
1/2 \tau_x^*(\B p) = i Z/2 \tau_x + i Z |{\rm Im} \Sigma_{\rm int}(\B p,\zeta_\B p)|
\eea
where we introduced the absolute value of the imaginary part of the self energy, which changes sign at $\omega=0$. Using the above we can rewrite the exciton propagator for low momenta as: 
\bea
G_x(p) \simeq \frac{Z}{\omega - \zeta_\B p + i \tau^*_x(\B p) \sgn(\omega)} \equiv Z \bar{G}_x(p)
\eea
which defines the projected operator $\bar{G}_x(p)$.  

In the main text we sometimes need to evaluate the partial derivatives of $\Sigma(\B p)$ on shell,i.e. at  $p=(\B p,\zeta_\B p)$. From Eq.~(\ref{ZZ}) we immediately have: 
\bea
\left. \partial_\omega \Sigma_{\rm int}(\B p,\omega)\right|_{\omega=\zeta_\B p} \simeq 1-Z^{-1}
\eea
To evaluate the momentum derivative we derivate Eq.~(\ref{dispPolaron}) to obtain: 
\bea
\frac{\di \zeta_\B p}{\di \B p}= \frac{\B p}{m_x^*} = \frac{\di \omega_\B p}{\di \B p} + \frac{\di \left({\rm Re} \Sigma_{\rm int}(\B p,\zeta_\B p)\right)}{\di \B p} = \frac{\B p}{m_x} - \partial_\B p {\rm Re} \Sigma_{\rm int}(\B p,\zeta_\B p) + \frac{\di \zeta_\B p}{\di \B p} \partial_\omega \left. {\rm Re} \Sigma_{\rm int}(\B p,\omega) \right|_{\omega=\zeta_\B p} 
\eea
From the above we immediately obtain: 
\bea
\partial_\B p {\rm Re} \Sigma_{\rm int}(\B p,\zeta_\B p) = Z^{-1}\frac{\B p}{m_x^*}-\frac{\B p}{m_x}
\eea

It is useful to obtain an explicit expression for the polaron dispersion $\zeta_\B p$ with respect to the value (and derivatives) of the self energy at the Fermi surface $\B p=\B p_F$, by expanding the self energy in a Taylor series with respect to these points: 
\bea
{\rm Re} \Sigma_{\rm int}(\B p,\omega) \simeq {\rm Re} \Sigma_{\rm int}(\B p_F,0) + \B p \partial_\B p\left. {\rm Re} \Sigma_{\rm int}(\B p,0)\right|_{\B p=\B p_F}+\frac{\B p^2}{2} \partial^2_\B p\left. {\rm Re} \Sigma_{\rm int}(\B p,0)\right|_{\B p=\B p_F}+ \omega \partial_\omega\left. {\rm Re} \Sigma_{\rm int}(\B p_F,0)\right|_{\omega=0}
\eea
The pole $\omega$ of Eq.~(\ref{A1}), which yields $\zeta_\B p$ therefore satisfies the equation:
\bea
\omega-\omega_\B p -{\rm Re} \Sigma_{\rm int}(\B p_F,0) - \B p \partial_\B p\left. {\rm Re} \Sigma_{\rm int}(\B p,0)\right|_{\B p=\B p_F}-\frac{\B p^2}{2} \partial^2_\B p\left. {\rm Re} \Sigma_{\rm int}(\B p,0)\right|_{\B p=\B p_F}- \omega \partial_\omega \left. {\rm Re} \Sigma_{\rm int}(\B p_F,0)\right|_{\omega=0} =0 
\eea
which can be solved to obtain: 
\bea
\zeta_\B p = Z\left( \omega_\B p+{\rm Re} \Sigma_{\rm int}(\B p_F,0)+ \frac{\B p^2}{2 m_x} - \B p \partial_\B p\left. {\rm Re} \Sigma_{\rm int}(\B p,0)\right|_{\B p=\B p_F}-\frac{\B p^2}{2} \partial^2_\B p\left. {\rm Re} \Sigma_{\rm int}(\B p,0)\right|_{\B p=\B p_F}\right)
\eea
where we introduced $\zeta_\B 0$. From the above it is clear that the renormalized mass of the polaron can be written as: 
\bea
\frac{1}{m_x^*} = Z\left(\frac{1}{m_x}+\partial^2_\B p\left. {\rm Re} \Sigma_{\rm int}(\B p,0)\right|_{\B p=\B p_F} \right)
\eea

\section{Polaron dispersion for a drifting Fermi sea}\label{app:polaron_dispersion}
Here, we determine the polaron dispersion, when the electrons are drifting at velocity $\B v_e$. 
For simplicity, we restrict ourselves to the case of zero polaron density, $\mu^*_x=0$. Moreover, we focus on sufficiently low energies (i.e., below the trion energy) such that we can represent the exciton Green's function by the effective polaron Green's function
\begin{align}
\bar{G}^R_x(p) = \frac{1}{\omega-\zeta_\B p+iZ/2\tau_x -i Z{\rm Im}\Sigma_{\rm int}(p)} ,\label{Gatt_app}
\end{align}
where $\zeta_\B p=\B p^2/2m_x^*$ is the polaron dispersion at zero density, when the electron Fermi sea is at rest.
According to Eq.~(\ref{eq20}), we can write the self energy in the absence of a drift as
\begin{align}
\Sigma^{(0)}_{\rm int}(\omega,{\bf p})&=   \int\frac{d\B k}{(2\pi)^2} n_F(\xi_{\bf k})
T^{(0)}(\omega+\xi_\B k,{\bf k}+{\bf p}).
\end{align}
The T matrix in Eq.~(\ref{eq21}) can be expressed as
\begin{align}
T^{-1}(\omega,{\bf p})
 &=V^{-1}+\int \frac{d\B q}{(2\pi)^2} \frac{n_F(\xi_{\bf q})-1}{\omega-\xi_\B q-\zeta_{\bf p-q}+iZ/2\tau_x -i Z{\rm Im}\Sigma_{\rm int}(\omega-\xi_\B q,\B p-\B q)},
\end{align}
where we have substituted the exciton Green's function by Eq.~(\ref{Gatt_app}). This is justified because in the domain of integration the energy argument of $G_x$ always remains smaller than $\omega$, and hence Eq.~(\ref{Gatt_app}) is a good approximation.
We can now simply introduce a drift velocity of the Fermi surface of electrons by shifting the distribution function $n_F(\xi_{k})\to n_F(\xi_{{\bf k-A}})$ with ${\bf A}={\bf v}_em_e$. We denote the self energy with a shifted Fermi surface as $\tilde \Sigma$ and find
\begin{align}
\tilde \Sigma^{(0)}_{\rm int}(\omega,{\bf p},{\bf A})&=   \int\frac{d\B k}{(2\pi)^2} n_F(\xi_{\bf k-A})
\tilde T(\omega+\xi_\B k,{\bf k}+{\bf p},{\bf A})\label{Sigmatilde_def}.
\end{align}
The self-consistent T matrix needs to be changed accordingly
\begin{align}
 \tilde T^{-1}(\omega,{\bf p},{\bf A})
 &=V^{-1}+\int \frac{d\B q}{(2\pi)^2} \frac{n_F(\xi_{\bf q-A})-1}{\omega-\xi_\B q-\tilde \zeta_{\bf p-q}+iZ/2\tau_x -i Z{\rm Im}\tilde{\Sigma}_{\rm int}(\omega-\xi_\B q,\B p-\B q,\B A)},\label{Ttilde}
\end{align}
where we have defined the polaron dispersion in the presence of a Fermi sea $\tilde \zeta_\B p$ that we seek to obtain.
Shifting the variable of integration the self energy reads
\begin{align}
\tilde \Sigma^{(0)}_{\rm int}(\omega,{\bf p},{\bf A})&=   \int\frac{d\B k}{(2\pi)^2} n_F(\xi_{\bf k})
\tilde T(\omega+\xi_{\bf k+A},{\bf k+A}+{\bf p},{\bf A})\label{sigtilde}
\end{align}
with
\begin{align}
 \tilde T^{-1}&(\omega+\xi_{\bf k+A},{\bf k+A}+{\bf p},{\bf A})
 =V^{-1}\notag\\
& +\int \frac{d\B q}{(2\pi)^2} \frac{n_F(\xi_{\bf q})-1}{\omega+\xi_{\bf k+A}-\xi_{\bf q+A}-\tilde \zeta_{\bf k+p-q}+iZ/2\tau_x -i Z{\rm Im}\tilde{\Sigma}_{\rm int}(\omega+\xi_{\bf k+A}-\xi_{\B q+\B A},{\bf k}+\B p-\B q,\B A)}.\label{Ttilde_shift}
\end{align}
We make the following general ansatz for the new polaron dispersion
\begin{align}
 \tilde \zeta_\B p=\frac{(\B p+\delta \B p)^2}{2m_x^*}+\delta E,\label{shifted_polaron_dispersion}
\end{align}
where $\delta \B p$ and $\delta E$ are constants to be determined. It is straightforward to rewrite the real part of the denominator of Eq.~(\ref{Ttilde_shift}) as
\begin{align}
\omega+\xi_{\bf k+A}- \xi_{\bf q+A}-\tilde \zeta_{\bf k+p-q}
&= \omega'+\xi_\B k -\xi_\B q- \zeta_{\bf k+p'-q},\label{denominator}
\end{align}
where we have introduced
\begin{align}
{\bf p}'=&{\bf p}+\delta \B p-{\bf A}m_{x}^*/m_e.\\
 \omega'=&\omega-\frac{(\B p+\delta \B p)\B A}{m_e}+\frac{A^2m_x^*}{2m_e^2}-\delta E.
\end{align}
Notice that the right-hand side of Eq.~(\ref{denominator}) involves the bare polaron dispersion at zero electron drift velocity.
Using this relation, we can express the T matrix in the presence of an electron drift by the bare T matrix at shifted energy and momentum arguments,
 $\tilde T(\omega+\xi_{\bf k+A},{\bf k+A}+{\bf p},{\bf A})= T(\omega'+\xi_{\bf k},{\bf k}+{\bf p}')$.
Similarly the self energy can be expressed as
\begin{align}
 \tilde \Sigma^{(0)}_{\rm int}(\omega,{\bf p},{\bf A})= \Sigma^{(0)}_{\rm int}(\omega',{\bf p}')\label{Sigmatilde}.
\end{align}
Here we have used the relation
\begin{align}
 {\rm Im}\tilde{\Sigma}_{\rm int}(\omega+\xi_{\bf k+A}-\xi_{\B q+\B A},{\bf k}+\B p-\B q,\B A)
 ={\rm Im}\Sigma_{\rm int}(\omega+\xi_{\bf k}-\xi_{\B q},{\bf k}+\B p-\B q),
\end{align}
which can be verified straightforwardly using Eqs.~(\ref{sigtilde}) and (\ref{Ttilde_shift}).
We can now determine the polaron dispersion in Eq.~(\ref{shifted_polaron_dispersion}) from the pole of the Green's function
\begin{align}
\Bigl[ \omega-\frac{p^2}{2m_x}+\mu_x-{\rm Re}\tilde\Sigma(\omega,{\bf p},{\bf A})\Bigr]_{\omega=\tilde \zeta_\B p}=0.\label{Green_pole}
\end{align}
Near the polaron pole we can write
\begin{align}
{\rm Re} \tilde \Sigma^{(0)}_{\rm int}(\omega,{\bf p},{\bf A}) ={\rm Re}\Sigma^{(0)}_{\rm int}(\omega',{\bf p}')\simeq(1-Z^{-1})\omega'+Z^{-1}\zeta_{p'}  -\frac{p'^2}{2m_x}+\mu_x.
\end{align}
Substituting $\B A=m_e \B v_e$ we find after some straightforward manipulations
\begin{align}
 \delta \B p=&{\bf v}_e(m_x^*- m_x)\\
 \delta E=&-\frac{v_e^2(m_x^*-m_x)}{2}
\end{align}
and hence the polaron dispersion when the electrons are drifting at a constant velocity reads
\begin{align}
 \tilde \zeta_\B p=\frac{(\B p+{\bf v}_e(m_x^*- m_x))^2}{2m_x^*}-\frac{v_e^2(m_x^*-m_x)}{2}.
\end{align}

 \section{Polaron transport using a variational approach}\label{sec:variational}
A complementary way to show the emergence of this force is using a variational approach. To do this we start from the following Hamiltonian that incorporates the interaction between excitons and electrons in the presence of an electric field. We will use the Coulomb gauge, such that the effect of the electric field is to shift the electron dispersion, and therefore preserve translational invariance. We obtain:
 \bea\label{ham}
 H(t)&=&\sum_{\B{k}} \xi_{\B{k}+m_e \B v_e(t)}   {c}_\B{k}^\dagger {c}_\B{k}+ \sum_\B{k} \omega_\B{k} {x}_\B{k}^\dagger {x}_\B{k}+\sum_{\B{k},\B{k}',\B{q}} \frac{V}{A} {x}_{\B{k}+\B{q}}^\dagger c^\dagger_{\B{k}'-\B{q}} {c}_{\B{k}'} {x}_\B{k}
 \eea
 where $c^\dagger$ is the electron creation operator, while $x^\dagger$  denotes the creation of an excitonic impurity. Furthermore, $\xi_\B k=\frac{\B k^2}{2 m_e}-\mu_e$ is the electron dispersion,  while $\omega_\B k=\frac{\B k^2}{2m_x}-\mu_x$ denotes the impurity dispersion. Notice that we investigate only the case of vanishing exciton density, i.e.  $\mu_x<0$ .  We remark that the total conjugate momentum $\hat {\B p}_T \equiv \sum_ \B k \B k \left( x^\dagger_ \B k x_\B k +c^\dagger_\B k c_\B k\right)$ is an integral of motion and we can simply replace it by its eigenvalue $\B p_T$.

 Strictly speaking the above Hamiltonian is valid only when disorder can be neglected and in this case $\B v_e(t)$ is the velocity acquired by electrons due to the acceleration by the electric field: $\B v_e(t)= \mathrm{Re} \frac{e}{i \Omega m_e} \B E e^{-i \Omega t}$. However, we can also include the effect of disorder on the electron system, heuristically, by assuming that the velocity $\B v_e(t)$ is the steady state electron velocity in the presence of an electric field and disorder (as calculated in Eq.~\ref{eq2}).

 It is instructive to first solve the problem in the absence of an electric field. To do this we introduce a Chevy-ansatz, which is completely equivalent to a non-self-consistent T-matrix approach:
 \bea
 \ket{\Psi_{\B{p}}} = a^\dagger_\B{p} \ket{0}=\left( \phi_{\B{p}} x^\dagger_\B{p} +\sum_{\B{k},\B{q}} \phi_{\B{p},\B{k},\B{q}} x^\dagger_{\B{p}+\B{q}-\B{k}} c^\dagger_\B{k} c_\B{q}  \right) \ket{0}\nonumber\\
 \eea
To obtain the ground state energy we have to minimize $\bra{ \Psi_{\B p}} H- E \ket{\Psi_{\B p}}$, where the energy $E$ is the Langrange multiplier ensuring the normalization of the wavefunction. The minimization will yield, the dispersion of the polarons $\zeta_\B p(t) =\zeta_\B 0+ \B p^2/(2 m_x^*)$ (see Ref.~\cite{chevy2006universal,sidler2017fermi} for details regarding the minimization procedure; the mass $m_x^*$ might not be the same as the $m_x^*$ obtained self-consistently, since this is a non-self-consistent derivation). 

Having solved the problem in the absence of the electric field, we now find a mapping from the instantenous eigenstates of the Hamiltonian in the presence of an electric field to the states in the absence of any field. To show this mapping we firtly go to a frame that is co-moving with the electrons with the unitary $U(t)=e^{i S(t)}$ where $S(t)=\B{r}_e(t) \hat{\B{p}}_{T}$, where $\B r_e(t)=\int_0^t \mathrm{d} t' \B v_e(t')$ . Since $U(t) x_{\B{k}} U^{\dagger}(t) = x_{\B k} e^{-i \B k \B r_e(t)}  $ , the conservation of total conjugate momentum implies that $U(t) H U^{\dagger}(t) = H$, so the only contribution to the Hamiltonian comes from the time dependence of the transformation $-i U(t) \frac{\partial U^\dagger(t)}{\partial t} = \frac{\partial S(t)}{\partial t}= \frac{e}{m_e} \B{A}(t) \hat {\B p}_T$. The new Hamiltonian becomes:
 \bea\label{Hamiltonian}
 H(t)&=&\sum_{\B{k}} \xi_\B k  {c}_\B{k}^\dagger {c}_\B{k}+ \sum_\B{k} \omega_{\B{k}-m_x \B{v}_e(t)} {x}_\B{k}^\dagger {x}_\B{k}+\sum_{\B{k},\B{k}',\B{q}} \frac{V}{A} {x}_{\B{k}+\B{q}}^\dagger e^\dagger_{\B{k}'-\B{q}} {e}_{\B{k}'} {x}_\B{k}.
 \eea
 where now the exciton dispersion is shifted by an amount $-m_x \B v_e(t)$. Assuming that the electric field is small enough, the system will remain in the many-body ground state, according to the adiabatic theorem. To determine the instantenous groundstate of  $H(t)$  we have to minimize $\bra{\Psi_\B{p}(t)} H (t) - E \ket{\Psi_{\B{p}}(t)}$, where the Chevy ansatz is given by:
 \bea
 \ket{\Psi_{\B{p}}(t)} = a^\dagger_\B{p} \ket{0}=\left( \phi_{\B{p}}(t) x^\dagger_\B{p} +\sum_{\B{k},\B{q}} \phi_{\B{p},\B{k},\B{q}}(t) x^\dagger_{\B{p}+\B{q}-\B{k}} c^\dagger_\B{k} c_\B{q}  \right) \ket{0}\nonumber\\.
 \eea
 One can check that, up to some irrelevant constants $\bra{\Psi_\B{p}(t)} H (t) \ket{\Psi_{\B{p}}(t)}=\bra{\Psi_{\B{p}-m_x \B v_e (t)}} H  \ket{\Psi_{\B{p}-m_x\B v_e(t)}}$ (one way to see this is to explicitly expand these terms and compare them) which illustrates the mapping to the states in the absence of an electric field.  From the above we immediately see that:
  \bea
  \tilde{\zeta}_\B{p}(t)=\zeta_{\B{p}-m_x \B v_e (t)}= \frac{\left(\B{p}-m_x \B v_e (t)\right.)^2}{2 m_x^*}- \mu_x^*,
  \eea
  where we used a tilde to denote the dispersion in the frame co-moving with the electrons. We remark that polarons acquire a backward velocity $-m_x/m_{x}^* \B v_e(t)$, which is smaller than the electron velocity due to the mass remormalization. Moving back to the lab frame, the dispersion will become:
 \bea
 \zeta_\B{p}(t)= \frac{\left(\B{p}+(m_x^*-m_x) \B v_e(t) \right.)^2}{2 m_x^*}-\mu_x^*
 \eea
 Assuming that the electric field is small enough the evolution is adiabatic and we can focus only on the attractive-polaron states and ignore all the other higher lying states and therefore we can write down an effective Hamiltonian in terms of attractive polarons:
 \bea
 H(t)=\sum_{\B p}\left( \frac{\left[\B{p}+(m_x^*-m_x) \B v_e(t)  \right]^2}{2 m_x^*} -\mu_x^* \right)a^\dagger_\B p a_\B p \nonumber\\
 \eea
which is similar to the Hamiltonian introduced in Sec.~(\ref{sec:low-energy}). 

\section{Polaron lifetime}\label{sec:scattering}
In this Appendix we investigate the residual interactions between polarons and the Fermi sea and calculate the corresponding scattering rates associated with these processes. We first calculate the scattering lifetime using Fermi's Golden Rule, since it is more transparent, but then we show that the same result can be obtained by explicitly evaluating ${\rm \Im}\Sigma(\B p, \xi_\B p)$. In the following we are interested in the case of only one polaron in the system with momentum $|\B p|\ll k_F$ and therefore we choose $\mu_x^*<0$. 

\subsection{Interaction between polarons and electrons}
The interaction between this polaron and an electron of momentum $\B k$ is given by:
\bea
U(\B p,\B k)=Z T(\B p +\B k,\zeta_\B p+\xi_\B k) \label{U}
\eea
where $Z$ is the quasiparticle weight of the polaron. 
Since, by assumption, the polaron has a small energy $\zeta_\B p \ll e_F$, the polaron will only be able to scatter electrons in a thin shell around the Fermi surface, and therefore, for our purposes we have $\B k\simeq \B k_F$ and therefore: 
\bea
U\simeq Z T(\B k_F,0) 
\eea
In the above $T$ denotes the self-consistent T-matrix. We know that the T-matrix has a a simple pole at the trion-resonance $\omega_T(k)$. Therefore, for energies in the vicinity of the trion resonance we can approximate the T-matrix by: 
\bea
T(\B k,\omega) \simeq \frac{C}{\omega-\omega_T(k)+i \gamma_T} 
\eea
where $\omega_T(\B k)$ is the trion energy measured from the exciton chemical potential $\mu_x$, and $\gamma_T$ denotes the lifetime of the trion-state. 

Since it is not immediately obvious how to calculate $C$ for the self-consistent T-matrix, we proceed to estimate it. Since the self-consistent T-matrix cannot be too much different than the non-self-consistent T-matrix $T_0$ (which we know how to calculate exactly) we use the analytical expression of $T_0$ to obtain an order of magnitude estimation for $T$. Considering, for simplicity the case $m_x=m_e$ we know that \cite{schmidt2012fermi}: 
\bea
T_0(0,\omega) = \frac{2}{\rho_e \left( \ln\left(\frac{\epsilon_T}{\omega+\mu_x}\right)+i \pi \right)} =\frac{2 \epsilon_T}{\rho_e}\frac{1}{\omega+\epsilon_T+\mu_x} + {\cal O}\left(\left(\frac{\omega+\epsilon_T+\mu_x}{\epsilon_T}\right)^2 \right)
\eea
which shows that $C\simeq 2 \epsilon_T / \rho_e$. Denoting the energy difference between the trion of momentum $\B k_F$ and the attractive polaron of momentum $\B p$ with $\Delta$ we can approximate the interaction $U$ as: 
\bea
U\simeq - \frac{2 Z\epsilon_T}{\rho_e \Delta} 
\eea

\subsection{Polaron lifetime using Fermi's Golden Rule}\label{sec:FGR}
According to Fermi's Golden rule, the scattering rate from a state of momentum $\B p$ is given by: 
\bea\label{scattering}
\Gamma = 2 \pi U^2 \sum_{|\B p'|<|\B p|,\B k} \delta(\zeta_{\B p} - \zeta_{\B p'}-\xi_{ \B k+\B p-\B p'}+\xi_{\B k} ) [n_F(\xi_\B k)-n_F(\xi_{ \B k+\B p-\B p'})]
\eea
where we used the fact that a polaron can scatter by creating electron-hole pairs in the Fermi sea. We can write the above more compactly by introducing the electron response function: 
\bea
\chi (\B q,\omega) \equiv -\sum_{\B k} \frac{n_F(\xi_\B k)-n_F(\xi_{\B k+\B q})}{\omega - \xi_{\B k+\B q} + \xi_{\B k} +i 0^+}
\eea
which allows us to rewrite $\Gamma$ as: 
\bea
\Gamma= 2 U^2 \sum_{|\B p'|<|\B p|} {\rm Im} \chi(\B p-\B p',\zeta_{\B p} - \zeta_{\B p'}) = 2 U^2 \sum_{|\B q|<|\B p| }{\rm Im} \chi(\B q,\zeta_{\B p} - \zeta_{\B p-\B q}) \theta(\zeta_{\B p} - \zeta_{\B p-\B q})
\eea
Since $|\B q|<|\B p|\ll k_F$ and therefore $\zeta_{\B p} - \zeta_{\B p-\B q}\ll \mu_e$ we can use the low frequency expansion: 
\bea
{\rm Im} \chi (\B q,\omega) \simeq 2 \rho_e \frac{\omega}{\mu_e} \frac{k_F}{|\B q|} \label{chiapprox}
\eea 
which in our case means that: 
\bea
\Gamma &\simeq& 4 \rho_e U^2 \int_{0}^p \di q q \int_{0}^{2 \pi}\di \phi {\rm Im} 2 \rho_e\frac{2 p q \cos \phi  -q^2 }{2 m_x^* \mu_e} \frac{k_F}{|\B q|}\theta(2 p q \cos \phi  -q^2)\\
& =& 4 \rho_e U^2 \frac{p^3 k_F}{2 m_x^* \mu_e}\int_{0}^1 \di q \int_{0}^{2 \pi} \di \phi \left(2 q \cos \phi - q^2\right)\theta(2 q \cos \phi-q^2)
\eea
The integral is easy to evaluate and is $\simeq 1$. Since $Z\simeq \sqrt{\mu_e/\epsilon_T}$ we finally obtain 
\bea\label{p3}
\Gamma \sim \frac{\epsilon_T}{\Delta^2} \frac{p^3 k_F}{m_em_x^*}.
\eea

\subsection{Polaron lifetime from self energy}\label{sec:polaron_lifetime_self_energy}
The same expression can be derived by explicitly evaluating the imaginary part of the self energy. To evaluate the imaginary part one can use the optical theorem. In our case it is just as simple to directly evaluate the imaginary part of the self energy. The polaron broadening due to interaction with the Fermi sea is related to the self energy by: 
\bea
\Gamma = Z {\rm Im}\Sigma(\B p,\zeta_\B p)
\eea
where: 
\bea
\Sigma(\B p,\omega)= \int \frac{\di^2 \B k}{(2 \pi)^2} n_F(\xi_{\B k}) T(\B p+\B k,\omega + \xi_{\B k})
\eea
where the $T$ matrix is given by:
\bea
T(\B p,\omega)^{-1} =V^{-1}- \int \frac{\di^2 \B k}{(2 \pi)^2} (1-n_F(\xi_\B k)) G^R_x(\B p-\B k,\omega-\xi_{\B k}) 
\eea
From the above we conclude that: 
\bea
\mathrm{Im} T(\B p,\omega)^{-1}&=&\int \frac{\di^2 \B k}{(2 \pi)^2} (1-n_F(\xi_\B k)) {\rm Im} G^A_x(\B p-\B k,\omega-\xi_{\B k}) \\
\mathrm{Im} T(\B p,\omega)&=&|T(\B p,\omega)|^2\mathrm{Im} T(\B p,\omega)^{-1}\simeq T(\B p,\omega)^2\mathrm{Im} T(\B p,\omega)^{-1}  \\
\mathrm{Im} \Sigma(\B p,\omega)&=&\int \frac{\di^2 \B k}{(2 \pi)^2} \frac{\di^2 \B k'}{(2 \pi)^2} T(\B p+\B k,\omega + \xi_{\B k})^2 (1-n_F(\xi_\B k'))n_F(\xi_{\B k}) {\rm Im}G^A_x(\B p+\B k-\B k',\omega+\xi_{\B k}-\xi_{\B k'}) \\
&=& \int \frac{\di^2 \B k}{(2 \pi)^2} \frac{\di^2 \B k'}{(2 \pi)^2} T(\B p+\B k,\omega + \xi_{\B k})^2 (1-n_F(\xi_\B k'))n_F(\xi_{\B k}) \pi Z \delta(\omega-\zeta_{\B p+\B k-\B k'}+\xi_{\B k}-\xi_{\B k'})
\eea
From the above we see that the polaron broadening will be: 
\bea
\Gamma = Z {\rm Im}\Sigma(\B p,\zeta_\B p) = Z^2 \pi \int \frac{\di^2 \B k}{(2 \pi)^2} \frac{\di^2 \B k'}{(2 \pi)^2} T(\B p+\B k,\zeta_\B p + \xi_{\B k})^2 (1-n_F(\xi_\B k'))n_F(\xi_{\B k}) \delta(\zeta_\B p-\zeta_{\B p+\B k-\B k'}+\xi_{\B k}-\xi_{\B k'})\nonumber\\
\eea
which, is can be directly related to Eq.~(\ref{scattering}) through the change of coordinates $\B k' \to -\B p' + \B k + \B p$ and using the approximation $T(\B p+\B k,\zeta_\B p + \xi_{\B k})\simeq T(\B k_F,0)$ which is valid for $|\B p|\ll k_F$ .

\subsection{Polaron lifetime at finite exciton density}\label{sec:lifetimeInMedium}
The above calculation was performed in the limit of vanishing exciton density, but we can use the same method to calculate the lifetime in the presence of a finite density of excitons. We therefore calculate the imaginary part of the interaction self energy for a polaron of frequency $\omega<0$ and momentum $|\B p|<|\B p_F|$. This corresponds to a hole of energy $-\omega$ and momentum $-\B p$, which can scatter into states with momenta $\B p'$ obeying the  condition $0\geq \zeta_{\B p'} \geq \omega $ by creating electron-hole pairs. 

Using similar arguments as in Sec.~(\ref{sec:FGR}) the scattering rate can be calculated using Fermi's Golden Rule: 
\bea
\Gamma (\B p,\omega) = 2 U^2 \sum_{\B p',|\B p'|<|\B p_F|}  {\rm Im} \chi(\B p-\B p',\omega - \zeta_{\B p'})\theta(\zeta_{\B p'}-\omega) \simeq \Gamma(\B p,\zeta_\B p) +  (\omega-\zeta_\B p) \partial_\omega \Gamma(\B p,\omega)_{\omega=\zeta_\B p}
\eea
where we expanded $\Gamma$ in a Taylor series in frequency around $\zeta_\B p$ and kept only the first term:
\bea
\Gamma(\B p,\zeta_\B p)&= &2 U^2 \sum_{\B p',|\B p'|<|\B p_F|}  {\rm Im} \chi(\B p-\B p',\zeta_\B p - \zeta_{\B p'})\theta(\zeta_{\B p'}-\zeta_{\B p})\\
\partial_\omega \Gamma(\B p,\omega)_{\omega=\zeta_\B p}&= &2 U^2  \partial_\omega \left[ \sum_{\B p',|\B p'|<|\B p_F|}  {\rm Im} \chi(\B p-\B p',\omega - \zeta_{\B p'})\theta(\zeta_{\B p'}-\omega) \right]_{\omega=\zeta_\B p}
\eea
Introducing $\B q= \B p-\B p'$ and using the low frequency and momentum expansion of the response function $\chi$, as given in Eq.~(\ref{chiapprox}) we obtain: 
\bea
\Gamma(\B p,\zeta_\B p)&\simeq & \frac{4 U^2  \rho_e k_F}{\mu_e }   \sum_{\B q,|\B p-\B q|<p_F}  \frac{\zeta_\B p-\zeta_{\B p-\B q }}{|\B q|} \theta(\zeta_{\B p-\B q}-\zeta_\B p)\\
&=&\frac{U^2  \rho_e k_F}{\mu_e \pi^2}  \int_{0}^{2 p_F} \di q q \int_0^{2 \pi} \di \phi   \frac{q^2-2 p q \cos \phi}{2 m_x^*}\frac{1}{q}\theta(q^2-2 p q \cos \phi) \theta(p_F^2+2 p q \cos \phi - p^2 - q^2) \\
&=&  \frac{U^2  \rho_e k_F}{\mu_e \pi^2}  \frac{p_F^3}{2 m_x^*} I_0(p/p_F) 
\eea
where we introduced the integral of order 1:
\bea
I_0(u)\equiv\int_0^2 \di q  \int_0^{2 \pi} \di \phi  (q^2-2 q u \cos \phi)\theta(q^2-2 q u \cos \phi)  \theta(1+2 u q \cos \phi - u^2 - q^2) 
\eea
In the above, one of the theta functions ensures that $\zeta_{\B p'}\geq \omega$ while the other theta function ensures that $|\B p'|<p_F$. It is useful to investigate the energy of hole quasiparticles with $-\omega=\zeta_\B p \ll \mu_x^*$. For these particles $p/p_F \simeq 1 - \omega/2 \mu_x^*$. Introducing $\delta= \omega/\mu_x^*$ one can easily check that $I_0(1-\delta/2) = C \delta^2 + {\cal O}(\delta^3)$. This proves that, in analogy to Fermi Liquid theory, the polaron excitations are well defined quasiparticles, with lifetime proportional to $\omega^2$.

The derivative of $\Gamma$ can be evaluated similarly using Eq.~(\ref{chiapprox}) : 
\bea
\partial_\omega \Gamma(\B p,\omega)_{\omega=\zeta_\B p} &=& \frac{4 U^2  \rho_e k_F}{\mu_e } \partial_\omega \left[   \sum_{\B q,|\B p-\B q|<p_F}  \frac{\omega-\zeta_{\B p-\B q }}{|\B q|} \theta(\zeta_{\B p-\B q}-\omega) \right]_{\omega=\zeta_\B p}\\
& =& \frac{4 U^2  \rho_e k_F}{\mu_e }  \left[   \sum_{\B q,|\B p-\B q|<p_F}  \frac{1}{|\B q|} \theta(\zeta_{\B p-\B q}-\zeta_\B p)- \frac{\zeta_\B p-\zeta_{\B p-\B q }}{|\B q|} \delta(\zeta_{\B p-\B q}-\zeta_\B p)\right] 
\eea
Similar to the evaluation of $\Gamma$ we can evaluate the above by turning the sum into integrals. We immediately see that the last term in the brackets vanishes and we are left with 
\bea
\partial_\omega \Gamma(\B p,\omega)_{\omega=\zeta_\B p} &=&\frac{U^2  \rho_e k_F}{\mu_e \pi^2} p_F I_1(p/p_F) 
\eea
where we introduced the integral of order 1:
\bea
I_1(u)\equiv\int_0^2 \di q  \int_0^{2 \pi} \di \phi  \theta(q^2-2 q u \cos \phi) \theta(1+2 u q \cos \phi - u^2 - q^2)
\eea

\section{\texorpdfstring{Relation between $\protect{\rm Im}\Sigma_{\rm int}$ and $\protect{\rm Im}G_x$}{}}\label{app:polaron_density}

In Sec.~\ref{sec:calculate_conductivity} of the main text, we encounter an expression of the type
\begin{align}
 \int dp A[G^R_x(p)] {\rm Im}\Sigma_{\rm int}(p)\theta(-\omega)
\end{align}
where $A$ is a functional of $G^R_x$. In this appendix, we show that this expression can alternatively be written as
\begin{align}
{\cal I}\equiv  \int dp A[G^R_x(p)] {\rm Im}\Sigma_{\rm int}(p)\theta(-\omega)=& \int \di p\di k A[G^R_x(p)] {\rm Im}G^A_x(k) w^{(0)}(k,p)\theta(-\epsilon)\notag\\
 =&\int \di p\di k\di k' A[G^R_x(p)] {\rm Im}G^A_x(k)
  G_e(k') G_e(p+k'-k)[T^R(p+k')]^2\theta(-\epsilon)\label{ImSigma_ImG}
\end{align}
where $A$ is a functional of $G^R_x$. 
We first derive an expression for the imaginary part of the self energy. Using $G_e(k)=G^R_e(k)+2i{\rm Im}G_e(k)\theta(-\epsilon)$ and a similar relation for the T matrix we can write the self energy as
\begin{align}
\Sigma_{\rm int}(p)\theta(-\omega)=&-i\theta(-\omega)\int \di k \bigr[G^R_e(k)T^R(k+p)+G^R_e(k)2i{\rm Im}T^A(k+p)\theta(-\omega-\epsilon)\notag\\
&+2i{\rm Im}G_e^A(k)\theta(-\epsilon)T^R(k+p)
-4{\rm Im}G_e^A(k){\rm Im}T^A(k+p)\theta(-\omega-\epsilon)\theta(-\epsilon)\bigl]\\
=&2\theta(-\omega)\int \di k \bigl\{{\rm Re}G_e(k){\rm Im}T^A(k+p)\theta(-\omega-\epsilon)
+{\rm Im}G_e^A(k){\rm Re}T(k+p)\theta(-\epsilon)\notag\\
&+i{\rm Im}G_e^A(k){\rm Im}T^A(k+p)[\theta(-\omega-\epsilon)-\theta(-\epsilon)]
\bigr\}
\end{align}
The imaginary part of the T matrix is
\begin{align}
 {\rm Im} T(k+p)=2|T(k+p)|^2\int dk' {\rm Im}G^A_e(k'){\rm Im}G^A_x(k+p-k')\theta(-\epsilon')\theta(\epsilon'-\omega-\epsilon).
\end{align}
With this we can express the imaginary part of the self energy as
\begin{align}
 {\rm Im}\Sigma_{\rm int}(p)\theta(-\omega)=&4\theta(-\omega)\int \di k \di k' |T(k+p)|^2 {\rm Im}G^A_e(k) {\rm Im} G^A_e(k'){\rm Im}G^A_x(p+k-k')\\
 &\times \theta(-\epsilon')[\theta(-\omega-\epsilon)-\theta(-\epsilon)] \theta(\epsilon'-\epsilon-\omega)\label{ImSigma}.
\end{align}
To prove Eq~(\ref{ImSigma_ImG}), we make the replacement $G_e(p+k'-k)=G_e^R(p+k'-k)+2i{\rm Im}G^A_e(p+k'-k)\theta(\epsilon-\omega-\epsilon')$. The first term is canceled by the integral over $\omega$ since all poles are located in the same complex half plane and hence after shifting $k'\to k'-p$, we find
\begin{align}
 {\cal I}
 =&2i\int \di p\di k\di k' A[G^R_x(p)] {\rm Im}G^A_x(k)
  G_e(k'-p)
 {\rm Im}G^A_e(k'-k)  [T^R(k')]^2\theta(-\epsilon)\theta(\epsilon-\epsilon')
\end{align}
The shift of variables conveniently allows us to repeat the same trick writing $ G_e(k'-p)= G^A_e(k'-p)-2i{\rm Im} G^A_e(k'-p)\theta(\epsilon'-\omega)$. The first term is again canceled by the $\omega$ integration and we find after returning to the original integration variables
\begin{align}
 {\cal I}
 =4\int \di p\di k\di k' A[G^R_x(p)] {\rm Im}G^A_x(k)
{\rm Im} G^A_e(k'){\rm Im}G^A_e(p+k'-k)  [T^R(k'+p)]^2\theta(-\epsilon)\theta(\epsilon') \theta(\epsilon-\epsilon'-\omega)
\end{align}
By comparison with Eq.~(\ref{ImSigma}) and using the simple identity
\begin{align}
 \theta(-\epsilon)\theta(\epsilon') \theta(\epsilon-\epsilon'-\omega)
 =\theta(-\epsilon')[\theta(-\omega-\epsilon)-\theta(-\epsilon)] \theta(\epsilon'-\epsilon-\omega)\theta(-\omega),
\end{align}
we arrive at
\begin{align}
  \int dp A[G^R_x(p)] {\rm Im}\Sigma^{(0)}_{\rm int}(p)\theta(-\omega)
=\int \di p\di k A[G^R_x(p)] {\rm Im}G^A_x(k) w^{(0)}(k,p)\theta(-\epsilon).\label{relation_Sigma_W} 
\end{align}

\subsection{Exciton density}
We now show that the integral $ -i\int dp G_x(p)$ over the exciton Green's function indeed yields the correct polaron density
\bea
n_x\equiv \langle \sum_{\B p} x_\B p^\dagger(t) x_\B p(t)  \rangle = \sum_p n_x(\B p) = \sum_\B p \langle T[x_\B p(t) x_\B p^\dagger(t+0^+) ] \rangle = -i\sum_\B p \int \frac{\di \omega}{2 \pi} G_x ( \B p , \omega) e^{i \omega 0^+}
\eea
where the infinitesimal $\omega 0^+$ originates from time-ordering and ensures that the contour is closed in the upper-half plane. Using the decomposition $G(p)=G^R(p)+2i{\rm Im}G^A(p)\theta(-\omega)$ we find
\begin{align}
 -i\int \di p G_x(p) =& 2\int \di p   \mathrm{Im} G_x^A(p) \theta(-\omega)
 = \int \di p   \frac{1}{\tau_x^*} |G_x^R(p)|^2 \theta(-\omega)
 +2\int \di p   \mathrm{Im} \Sigma^{(0)}_{\rm int}(p) \theta(-\omega)[G_x^R(p)]^2
\end{align}
where we have expanded the second term to lowest order in $n_x$. As $1/\tau_x^*$ is a small parameter, the first term is dominated by on-shell contributions $\omega\simeq \zeta_p$ and we can write
\begin{align}
  \frac{1}{\tau_x^*} |G_x^R(p)|^2 \theta(-\omega)&\simeq  2\pi Z\frac{1/2\tau_x^*}{1/2\tau_x^*+{\rm Im}\Sigma_{\rm int}(p)} \delta(\omega-\zeta_p) \theta(-\omega)
  \simeq  2\pi Z \theta(-\omega) \delta(\omega-\zeta_p),
\end{align}
where the last line follows from ${\rm Im}\Sigma_{\rm int}(p)\ll 1/\tau_x^*$ for onshell energies $\zeta_p\simeq \omega\leq 0$ consistent with the assumptions made in Sec.~\ref{sec:calculate_conductivity}. The second term can be replaced by Eq.~(\ref{relation_Sigma_W}), which yields
\begin{align}
 -i\int \di p G_x(p)
 = Z n_x
+ 2&\int \di p    {\rm Im}G^A_x(k) \theta(-\epsilon)
 W^{(0)}(k,p)[G_x^R(p)]^2,
\end{align}
where the density is $n_x=\mu_x^*\nu_x^*$. We have $ \int \di p W^{(0)}(k,p)[G_x^R(p)]^2=-Z\partial_\omega\Sigma(p)=1-Z$ and, hence,
\begin{align}
 -i\int \di p G_x(p) =&   n_x.
\end{align}

\section{Proof of Eq.~(\ref{G_product_integral}) in main text}
\label{app:proof_product}

We now demonstrate Eq.~(\ref{G_product_integral}) of the main text, which reads
\begin{align}
 \frac{\B k}{m_e} G_e(k)G_e(k+\Omega)
\simeq \frac{\left(i\tau_e\Omega \frac{\B k}{m_e}  \partial_{\epsilon} +\partial_{\B k} \right) G_e( k)}{1-i\tau_e\Omega }\label{G_product_integral_app}
\end{align}
to leading order in $\Omega$ and $1/\tau_e$, where $G_e=[\epsilon-\xi_\B k+i(1/2\tau_e){\rm sgn}\epsilon]^{-1}$.
We start by writing $G_e(k)=G_e^R(k)+2i{\rm Im}G_e^A(k)\theta(-\epsilon)$ and we obtain
\begin{align}
 G_e(k)G_e(k+\Omega)=&G^R_e(k)G^R_e(k+\Omega)+{\rm Re}G^R_e(k)2i{\rm Im}G^A_e(k+\Omega)\theta(-\epsilon-\Omega)\notag\\
 &+{\rm Re}G^R_e(k+\Omega)2i{\rm Im}G^A_e(k)\theta(-\epsilon)+2{\rm Im}G^A_e(k){\rm Im}G^A_e(k+\Omega)[\theta(-\epsilon)-\theta(-\epsilon-\Omega)]\label{G_squared}
\end{align}
We are interested in a result to zeroth order in $\Omega$ and $1/\tau_e$. At this order, all but the first term vanish everywhere except when $\epsilon=\xi_\B k$. Our strategy is to expand these expressions in terms of $\delta$ functions around the resonance, for instance, we write
\begin{align}
  {\rm Re} G^R_e(k){\rm Im}G^A_e(k+\Omega)= \alpha_0\delta(\epsilon-\xi_\B k)+\alpha_1\delta'(\epsilon-\xi_\B k)+\alpha_2\delta''(\epsilon-\xi_\B k)+\ldots
\end{align}
We can obtain the coefficients from the following integrals
\begin{align}
\alpha_0=&\int d\epsilon {\rm Re} G^R_e(k){\rm Im}G^A_e(k+\Omega)=\frac{-\pi\Omega}{\tau_e^{-2}+\Omega^2},\\
\alpha_1=&-\int d\epsilon (\epsilon-\xi_\B k){\rm Re} G^R_e(k){\rm Im}G^A_e(k+\Omega)=-\frac{\pi[2(2\tau_e)^{-2}+\Omega^2]}{\tau_e^{-2}+\Omega^2},\\
\alpha_2=&\int d\epsilon (\epsilon-\xi_\B k)^2{\rm Re} G^R_e(k){\rm Im}G^A_e(k+\Omega)=-\frac{\pi \Omega[3 (2\tau_e)^{-2}+\Omega^2]}{\tau_e^{-2}+\Omega^2}.
\end{align}
The second order is already linear in the small parameters and can hence be ignored. The expansion then reads
\begin{align}
 {\rm Re} G^R_e(k){\rm Im}G^A_e(k+\Omega)\simeq \frac{\pi}{\tau_e^{-2}+\Omega^2}\Bigl[-\Omega\delta(\epsilon-\xi_\B k)-\Bigl(\frac{1}{2\tau_e^2}+\Omega^2\Bigr)\delta'(\epsilon-\xi_\B k)\Bigr].
\end{align}
Similar considerations yield
\begin{align}
 {\rm Re} G^R_e(k+\Omega){\rm Im}G^A_e(k)\simeq &\frac{\pi}{\tau_e^{-2}+\Omega^2}\Bigl[\Omega\delta(\epsilon-\xi_\B k)-\frac{1}{2\tau^2}\delta'(\epsilon-\xi_\B k)\Bigr]\\
 {\rm Im}G^A_e(k){\rm Im}G^A_e(k+\Omega)\simeq &\frac{\pi}{\tau_e^{-2}+\Omega^2}\Bigl[\frac{1}{\tau_e}\delta(\epsilon-\xi_\B k)+\frac{\Omega}{2\tau_e}\delta'(\epsilon-\xi_\B k)\Bigr]
\end{align}
Substituting these relations in Eq.~(\ref{G_squared}) we find
\begin{align}
 G_e(k)G_e(k+\Omega)\simeq&G^R_e(k)^2
 -2i\pi \delta'(\epsilon-\xi_\B k)\theta(-\epsilon)+\frac{2\pi\Omega\tau_e}{1-i\Omega\tau_e}\delta(\epsilon-\xi_\B k)\delta(\epsilon)
\end{align}
where we have expanded the step function $\theta(-\epsilon-\Omega)\simeq \theta(-\epsilon)-\Omega\delta(\epsilon)$. The right-hand side of Eq.~(\ref{G_product_integral_app}) can be evaluated explicitly using  $G_e(k)=G_e^R(k)+2\pi i\delta(\epsilon-\xi_\B k)\theta(-\epsilon)$. We obtain
\begin{align}
 \left(i\tau_e\Omega \frac{\B k}{m_e}  \partial_{\epsilon} +\partial_{\B k} \right) G_e( k)
 = (1-i\tau_e\Omega) \frac{\B k}{m_e}\Bigl\{G_e^R(k)^2-2\pi i\delta'(\epsilon-\xi_\B k)\theta(-\epsilon)
 +\frac{2\pi\tau_e\Omega}{1-i\Omega\tau_e} \delta(\epsilon-\xi_\B k)\delta(\epsilon)
 \Bigr\}
\end{align}
which concludes the proof.
\end{widetext}

\bibliographystyle{unsrt}
\bibliography{references}

\end{document}